\documentclass[a4paper,aps,showpacs]{revtex4}

\usepackage{graphics}
\usepackage{amssymb}
\usepackage{amsmath}

\usepackage{soul}
\usepackage{color}
\usepackage{graphics}
\usepackage{amsmath,amssymb,amsfonts}
\usepackage{amssymb} 
\usepackage{bm}
\usepackage{latexsym}
\usepackage{ulem}

\begin{document}
%%%%%%%%%%%%%%%%%%%%%%%%%%%%%%%%%%%%%%%%%%

%%%%%%%%%%%%%%%%%%%%%%%%%%%%%%%%%%%%%%%%%%%%%%%%%%%%%%%%%%%%%%%%%%%%%%
% Full title of the paper
\title{Noether Symmetry Analysis of the Klein--Gordon and Wave Equations in Bianchi I Spacetime}

\author{Ugur Camci}

\email{ucamci@rwu.edu,ugurcamci@gmail.com}

\bigskip

\affiliation{Department of Chemistry and Physics, Roger Williams University, Bristol, Rhode Island 02809, USA}

\bigskip

\date{\today}

%%%%%%%%%%%%%%%%%%%%%%%%%%%%%%%%%%%%%%%%%%%%%%%%%%%%%%%%%%%%%%%%%%%%%%%%%%%

\begin{abstract}
We investigate the Noether symmetries of the Klein--Gordon Lagrangian for Bianchi I spacetime. This is accomplished using a set of new Noether symmetry relations for the Klein--Gordon Lagrangian of Bianchi I spacetime, which reduces to the wave equation in a special case. A detailed Noether symmetry analysis of the Klein--Gordon and the wave equations for Bianchi I spacetime is presented, and the corresponding conservation laws are derived.
\end{abstract}

% \pacs{04.30, 04.30.Nk, 04.50.+h, 98.70.Vc}

% Keywords
% \keyword{Noether symmetry; Bianchi I spacetime; Klein--Gordon equation }

\pacs{04.20.-q, 11.30.-j, 98.80.-k}

\maketitle

%%%%%%%%%%%%%%%%%%%%%%%%%%%%%%%%%%%%%%%%%%%%%%%%%%%%%%%%%%%%%%%%%%%%%%%%%
%1
\section{Introduction}
\label{intro}

A conformal Killing vector (CKV) ${\bf K}$ has to satisfy
\begin{equation}
    \pounds_{\bf K} g_{ij} = 2 \, \sigma (x^k) g_{ij} \, , \label{ckv-eq}
\end{equation}
where $g_{ij}$ is the metric tensor, $\pounds_{\bf K}$ is the Lie derivative operator along ${\bf K}$, and $\sigma (x^k)$ is a conformal factor. When $\sigma_{;ij} \neq 0$, the CKV field is said to be {\it proper} \cite{katzin}. The vector field ${\bf K}$ is called a special conformal Killing vector (SCKV) field if $\sigma_{;ij} = 0$; a homothetic Killing vector (HKV) field if $\sigma_{, \, i} = 0$, e.g., $\sigma$ is a constant on the manifold; and a Killing vector (KV) field if $\sigma =0$, which is also called the isometry of spacetime. The set of all CKVs (respectively SCKV, HKV, and KV) forms a finite-dimensional Lie algebra. The maximum dimensions of the CKV algebra on the manifold $M$ is {\it fifteen} if $M$ is conformally flat, and it is {\it seven} if the spacetime is not conformally flat. The physical features of differential equations, in terms of the conservation laws admitted by them, are directly associated with the Noether symmetries, which is facilitated using a Lagrangian of the corresponding dynamical system. As we show in the following section, there is a direct relation between the conformal symmetries and Noether symmetries.

If a Lagrangian $\mathcal{L}$ for a given dynamical system exhibits symmetry, this property is strongly related to Noether symmetries, which describe the physical characteristics of differential equations associated with a Lagrangian $\mathcal{L}$, in terms of the first integrals they possess \cite{stephani,ibrahim}. This relationship can be viewed from two different perspectives. First, one can take a strict Noether symmetry approach \cite{capo00,camci2007,camci2012a}, which results in $\pounds_{\bf X} \mathcal{L} = 0$, where $\pounds_{\bf X}$ is the Lie derivative operator along ${\bf X}$. On the other hand, one can employ the Noether symmetry approach with a gauge term \cite{feroze1,tsamparlis1,tsamparlis2,camci2012b,camci2014,ua2014,ua2015}, a generalization of the strict Noether symmetry approach where the Noether symmetry equation includes the gauge term.
Noether symmetries with a gauge term are equally valuable in addressing a variety of problems in physics and applied mathematics. In the following section, we will discuss the relationship between Noether symmetries with a gauge term of the Klein--Gordon Lagrangian and the geometric symmetries of spacetimes. The study of differential equations involving geometry is an active area of research.
Recent literature \cite{pt,ptm} has delved into the connection between geometrical structures and conserved quantities. Noether symmetries are directly linked to conserved quantities or conservation laws \cite{ibrahim}, which naturally emerge in a wide range of applications.

The cosmological principle assumes that the universe is homogeneous and isotropic at large scales, and the geometrical model that satisfies these properties is Friedmann--Lema\^{i}tre--Robertson--Walker (FLRW) spacetime. However, there has been a suggestion from studies of the cosmic microwave background (CMB) temperature anisotropies that the assumption of statistical isotropy is violated at the largest angular scales, leading to some intriguing anomalies \cite{Ellis}. In order to make predictions for the CMB anisotropies, one can explore cosmological models that are homogeneous but anisotropic, such as the Bianchi-type spacetimes, which encompass both isotropic and homogeneous FLRW models.

Bianchi classified all three-dimensional real Lie algebras and demonstrated that there are nine possible simply transitive groups of motions, denoted as $G_3$. These groups are generated using Killing vectors $K_{\alpha}$, where $\alpha = 1, 2, 3$, with structure constants $C^{\gamma}_{\alpha \beta}$, defined by $\left[ K_{\alpha}, K_{\beta} \right] = C^{\gamma}_{\alpha \beta} K_{\gamma}$. The structure constants $C^{\gamma}_{\alpha \beta}$ can be decomposed into irreducible parts, as follows:
$$C^{\gamma}_{\alpha \beta} = \epsilon_{\alpha \beta \sigma} N^{\sigma \gamma} + A_{\alpha} \delta^{\gamma}_{\beta} - A_{\beta} \delta^{\gamma}_{\alpha}$$
here $N^{\alpha \beta}$ is symmetric. It follows from the Jacobi identities that $N^{\alpha \beta} A_{\beta} = 0$. One can diagonalize $N_{\alpha \beta}$ without loss of generality, choosing $A_{\alpha}$ in the 1-direction when it is non-zero. The real Lie algebras can then be classified into nine types of spatially homogeneous Bianchi spacetimes, distinguished by the particular form of the structure constants $C^{\gamma}_{\alpha \beta}$, based on whether $A_{\alpha} = 0$ (Class A) or $A_{\alpha} \neq 0$ (Class B). This classification results in the Bianchi types, such as $I$, $II$, $VI_0$, $VII_0$, $VIII$, and $IX$ for Class A models, and $III$, $IV$, $V$, $VI_h$, and $VII_h$ for Class B models. It is important to note that the Bianchi models include FLRW models as special cases, with Bianchi I (flat), Bianchi V (open), and Bianchi IX (closed) representing the FLRW models.
In this study, we consider the Bianchi I metric as the background spacetime in the Klein--Gordon equation. The line element for Bianchi I spacetime can be written as follows \cite{kramer}:
\begin{equation}
ds^2 = - dt^2 + A(t)^2 dx^2 + B(t)^2 dy^2 + C(t)^2 dz^2 \, . \label{b1}
\end{equation}
The above spacetime yields a flat FLRW metric if $A = B = C$. For any form of the metric coefficients $A, B$, and $C$, it can easily be computed from Equation \eqref{ckv-eq} that the KVs for Bianchi I spacetime are ${\bf K}_1 = \partial_x, {\bf K}_2 = \partial_y$, and ${\bf K}_3 = \partial_z$. In this study, all vector fields are written with bold letters.

The rest of the paper is organized as follows: In the next section, \ref{sec:2}, we present an analysis of Noether symmetries with a gauge term for the Klein--Gordon Lagrangian in the context of the Bianchi I spacetime model. In Section \ref{sec:3}, we apply the Noether symmetry approach to the Bianchi I spacetime. In Section \ref{sec:4}, we study the field equations of the Bianchi I spacetime with an imperfect fluid. Conclusions and discussions are presented in the final section, Section \ref{sec:5}.

%2
\section{Klein--Gordon Lagrangian and Noether Symmetry Equations}
\label{sec:2}

The Klein--Gordon equation in the Riemannian space with metric $g_{ik}$ is a second-order partial differential equation of the form
\begin{equation}
\square \psi = G(x^k,\psi), \label{kg-eq}
\end{equation}
where $\square \psi = g^{ik} \psi_{;ik}$  and $\square$ refers to the de'Alembertian or Laplace operator defined by \mbox{$\square = \frac{1}{\sqrt{-g}} \frac{\partial}{\partial x^i} \left( \sqrt{-g} g^{ik} \frac{\partial}{\partial x^k} \right)$} in terms of the Riemannian space.
The Klein--Gordon equation given in \eqref{kg-eq} follows from  the first-order Lagrangian, which is called the Klein--Gordon Lagrangian,
\begin{equation}
\mathcal{L}= \sqrt{-g} \left( \frac{1}{2} g^{ij} \psi_i \psi_j - F(x^k,\psi) \right), \label{lagr}
\end{equation}
where $\psi_i (x^k) \equiv \frac{\partial \psi (x^k)}{ \partial x^i}$ and $F'(x^k,\psi)=-G(x^k,\psi)$. Here, the prime represents the derivative with respect to the scalar field $\psi$. The above Lagrangian reduces to the Lagrangian of the wave equation when $F = 0$.

The wave equation of spacetimes is one of the most important equations in physics, and it is common to study this equation in terms of the Lie and Noether symmetry generators they admit \cite{sj1,sj2,js2016,jamal1,abdul2017,abdul2021}. When symmetry generators exist, they play a crucial role in finding exact solutions. Jamal et al. \cite{jamal1} investigated the wave equation of Bianchi III spacetime, calculating and classifying Noether symmetries and constructing corresponding conservation laws. They also obtained reductions of the wave equation and identified some invariant solutions.
In the papers \cite{abdul2017,abdul2021}, the authors utilized the invariance and multiplier method \cite{olver, anco2002} to conduct conservation law classifications of the wave equation for the Bianchi I spacetime, using power law metric coefficients.

Noether symmetries of the Klein--Gordon equations for well-known spacetimes have been calculated and classified according to their symmetry generators \cite{ptm,jamal2,ptm2018,jkb2011, jamal3,p2023}.
In \cite{jamal2}, the symmetry analysis of the Klein--Gordon equation in de Sitter spacetime was classified, and the obtained symmetries were utilized to find exact solutions using quadratures. Paliathanasis et al. \cite{ptm2018} conducted a classification of Lie and Noether symmetries for the Klein--Gordon and wave equations in pp-wave spacetimes.
The symmetry properties and conservation laws of wave and Gordon-type equations in Milne and Bianchi III spacetimes were investigated in the papers \cite{jkb2011, jamal3}.
In \cite{ptm}, a geometric procedure was employed for the symmetry classification of the Klein--Gordon equation in Bianchi I spacetime, connecting the Noether symmetries with a gauge function of the Klein--Gordon Lagrangian to the conformal symmetries of the metric tensor. Their study extended the results of Bozhkov and Freire \cite{bozhkov} for the Klein--Gordon equation with a constant potential, i.e., $V(x^k) = V_0$.
More recently, Paliathanasis \cite{p2023} considered the Klein--Gordon equations for the conformal forms of Bianchi I, Bianchi III, and Bianchi V spacetimes, deriving closed-form expressions for potential functions that admit Lie and Noether symmetries of the Klein--Gordon equations.
In this work, we aim to derive Noether symmetries with a gauge term for the dynamical Lagrangian $\mathcal{L}$ of the Klein--Gordon equation within the background of Bianchi I spacetimes. To obtain the Klein--Gordon equation in this spacetime, we construct a Lagrangian model. Using the obtained Lagrangian for the Klein--Gordon equation in Bianchi I spacetime, we calculate and classify Noether symmetry generators with gauge terms. Furthermore, we identify conservation laws provided by the Lagrangian for representing the \mbox{Klein--Gordon equation}.

In a general program of research into the problem of integrating the classical and quantum equations of motion of a test particle in external fields of different nature in spaces with symmetry following the sets of Killing fields, Obukhov found all admissible electromagnetic fields for the case, when the groups of motions $G_3$ act simply transitively on the hypersurfaces of spacetime $V_4$ \cite{obukhov1,obukhov2,obukhov3,obukhov4}. In \cite{obukhov1}, he found all external electromagnetic fields in which the Klein--Gordon--Fock equation admits the first-order symmetry operators and completed the classification of admissible electromagnetic fields in which the Hamilton--Jacobi and Klein--Gordon--Fock equations admit algebras of motion integrals that are isomorphic to the algebras of operators of the $r$-parametric groups of motions, $G_r$, of spacetime manifolds if $r \leq 4$ \cite{obukhov2}. In the paper \cite{obukhov3}, the case when the groups $G_4$ act on $V_3$ was considered. The remaining case in the latter article, when the groups $G_4$ act simply transitively on the space $V_4$, was studied in the paper \cite{obukhov4}.

The Noether symmetry generator for the Klein--Gordon Lagrangian \eqref{lagr} is
\begin{eqnarray}
& & {\bf X} = \xi^i (x^k, \psi) \frac{\partial}{\partial x^i} + \Phi (x^k, \psi) \frac{\partial}{\partial \psi} \nonumber
\end{eqnarray}
if there a gauge function $f^i (x^k, \psi)$ exists and the Noether symmetry condition
\begin{equation}
{\bf X}^{[1]} \mathcal{L}+ \mathcal{L} ( D_i \xi^i )= D_i f^i \label{ngs-cond-1}
\end{equation}
is satisfied, where $$D_i = \frac{\partial}{\partial x^i} + \psi_i \frac{\partial}{\partial \psi} + \psi_{ik} \frac{\partial}{\partial \psi_k} + ...  $$ is the total derivative operator and ${\bf X}^{[1]}$ is the first prolongation of Noether symmetry generator ${\bf X}$, i.e.,
\begin{eqnarray}
& & {\bf X}^{[1]} = {\bf X} + {\zeta}_i (x^k, \psi, \psi_k) \frac{\partial}{\partial \psi_i} \label{first-pro}
\end{eqnarray}
where ${\zeta}_i (x^k, \psi, \psi_k) = D_i \Phi - \psi_j D_i \xi^j$.
The corresponding Noether flow $T^i$ is defined by \mbox{the expression}
\begin{equation}
T^i = - \xi^i \mathcal{L} + \left( \xi^k \psi_k - \Phi \right) \frac{\partial \mathcal{L}}{\partial \psi_i} + f^i, \label{c-vector}
\end{equation}
which is called the {\it conserved vector} ${\bf T} = (T^1,...,T^n)$ , where $i=1, ..., n$, or if $ n=1$, then it is called the {\it conserved quantity}. The Noether flow \eqref{c-vector} satisfies the local conservation law
\begin{equation}
D_i T^i = 0. \label{c-eq}
\end{equation}

It is crucial to discover conservation laws when studying physical systems. For studying differential equations, conservation laws are useful for integrability, linearization, analyzing solutions, and understanding constants of motion. Noether's theorem enables the derivation of all local conservation laws for a (system of) differential equation(s) derived from a Lagrangian.
This method helps resolve the problem of calculating conservation laws for given differential equation(s), as Noether's theorem provides a formula utilizing symmetries of the action to derive these local conservation laws. One can find a detailed discussion about the physical significance of Noether symmetries in \cite{au2022}.

For the Klein--Gordon Lagrangian \eqref{lagr}, we obtain the first prolongation of the Noether symmetry generator ${\bf X}$ as
\begin{eqnarray}
 & & {\bf X}^{[1]} \mathcal{L} = \sqrt{-g} \Big{\{} -\Phi F_{,\psi} - F \xi^i \Gamma^k_{\,\, ik} -F_{,i} \xi^i + \frac{1}{2} \left[  \pounds_{\xi} g^{mn} + \left( \xi^i \Gamma^k_{\,\, ik} + 2 \Phi_{,\psi} \right) g^{mn} \right] \psi_m \psi_n \nonumber \\& & \qquad \qquad \quad  \qquad  + g^{im}\left( \Phi_{,i} \psi_m - \xi^j_{,\psi} \psi_m \psi_i \psi_j \right) \Big{\}} \, , \label{1st-pro}
\end{eqnarray}
where $\pounds_{\xi}$ is the Lie derivative operator along ${\xi} = \xi^k \partial / \partial x^k$. Putting \eqref{1st-pro} into \eqref{ngs-cond-1} together with $D_i \xi^i = \xi^i_{,i} + \xi^i_{,\psi} \psi_i$ and $D_i f^i = f^i_{,i} + f^i_{,\psi} \psi_i$, the Noether symmetry condition \eqref{ngs-cond-1} gives rise to
\begin{eqnarray}
& & \xi^i_{,\psi} = 0, \,\,\, \sqrt{-g} g^{ij} \Phi_{,j} - f^i_{,\psi} = 0,  \label{ngs-cond-2-1} \\& & \qquad \pounds_{\xi} g_{ij} =  \left( \xi^k_{;k} + 2 \, \Phi_{,\psi} \right) g_{ij},  \label{ngs-cond-2-2} \\ & & \sqrt{-g} \left( \pounds_{\xi} F  + F \xi^i_{\, ;i} + \Phi F_{,\psi} \right) + f^i_{,i} = 0.  \label{ngs-cond-2-3}
\end{eqnarray}
Thus, we find the geometrical form of Noether symmetry equation \eqref{ngs-cond-1} in terms of Lie derivatives of the metric tensor. Here, Equation \eqref{ngs-cond-2-1} yields $\xi^i = \xi^i (x^k)$ and \linebreak $f^i = \sqrt{-g} g^{ij} \int{ \Phi_{,j} d\psi} + A^i (x^k)$, where $A^i (x^k)$ is an integration function. If $\xi^i$ is a CKV with conformal factor $\sigma (x^k)$, then Equations \eqref{ckv-eq} and \eqref{ngs-cond-2-2} imply $\xi^i_{;i} = 2 \, \sigma - 2 \, \Phi_{,\psi}$. Finally, the Noether symmetry condition \eqref{ngs-cond-2-3} becomes
\begin{equation}
  F_{,i} \xi^i + 2 \, F \, ( \sigma - \Phi_{,\psi} ) + \Phi \, F_{,\psi} + \int{(\square\, \Phi) d\psi} +  \frac{A^i_{,i}}{\sqrt{-g}} = 0 \, . \label{cond-potential-1}
\end{equation}
For the Bianchi I spacetime \eqref{b1}, the Klein--Gordon Lagrangian \eqref{lagr} has the form
\begin{eqnarray}
& & \mathcal{L}=  -\frac{A B C}{2} \psi^2_t + \frac{B C}{2 A} \psi^2_{x} + \frac{A C}{2 B} \psi^2_y   + \frac{A B}{2 C} \psi^2_z  - A B C\,  F(t,x,y,z,\psi) \, , \label{lagr-2}
\end{eqnarray}
which is the the Lagrangian of the wave equation when $F=0$.
The Klein--Gordon \mbox{Equation \eqref{kg-eq}} is obtained through variation of this Lagrangian with respect to the scalar field $\psi$, as the following\vspace{-6pt}:
\begin{eqnarray}
& & \frac{1}{A^2} \psi_{x x} + \frac{1}{B^2} \psi_{yy} + \frac{1}{C^2} \psi_{zz} - \psi_{tt} - \frac{(A B C)^{^{\centerdot}} }{A B C} \psi_t = G(t,x,y,z,\psi), \label{kg-eq-2}
\end{eqnarray}
where $G(t,x,y,z,\psi) = -\frac{\partial F}{\partial \psi}$ and the dot represents the derivative with respect to time $t$.

Let us consider the Noether symmetry generator for the Klein--Gordon Lagrangian \eqref{lagr-2} as follows:
\begin{equation}
{\bf X} = \xi^0 \frac{\partial}{\partial t} + \xi^1
\frac{\partial}{\partial x} + \xi^2 \frac{\partial}{\partial y} +
\xi^3 \frac{\partial}{\partial z} + \Phi \frac{\partial}{\partial \psi}, \label{vecf}
\end{equation}
where the components of ${\xi} = ( \xi^0, \xi^1, \xi^2, \xi^3 )$ and $\Phi$ are dependent on $t,x,y,z$ and $\psi$. Now, we seek the dependent variables $\xi^0, \xi^1, \xi^2, \xi^3, \Phi$ that will be solved from the geometrical Noether symmetry conditions \eqref{ngs-cond-2-1}--\eqref{ngs-cond-2-3}, in order that the Lagrangian \eqref{lagr-2} would admit any Noether symmetry.
For the Bianchi I spacetime \eqref{b1}, the geometrical Noether symmetry conditions \eqref{ngs-cond-2-1}--\eqref{ngs-cond-2-3} yield {\it 19} PDEs:
\begin{eqnarray}
& & \xi^0_{,\psi} = 0, \quad \xi^1_{,\psi} = 0, \quad \xi^2_{,\psi} = 0, \quad \xi^3_{,\psi} = 0, \quad A B C \, \Phi_{,t} + f^0_{,\psi} = 0, \quad \frac{B C}{A}\, \Phi_{,x} - f^1_{,\psi}  = 0, \nonumber \\& & \frac{A C}{B} \Phi_{,y} - f^2_{,\psi} = 0, \quad \frac{A B}{C} \Phi_{,z} -  f^3_{,\psi} = 0, \quad  A^2 \xi^1_{,t} - \xi^0_{,x} = 0, \quad B^2 \xi^2_{,t} - \xi^0_{,y} = 0, \nonumber \\& &  C^2 \xi^3_{,t} - \xi^0_{,z} = 0, \quad A^2 \xi^1_{,y} + B^2 \xi^2_{,x} = 0, \quad  A^2 \xi^1_{,z} + C^2 \xi^3_{,x} = 0, \quad  B^2 \xi^2_{,z} + C^2 \xi^3_{,y} =0, \nonumber  \\& & -\xi^0_{,t} + \xi^1_{,x} + \xi^2_{,y} + \xi^3_{,z} + \left( \frac{\dot{A}}{A} + \frac{\dot{B}}{B} + \frac{\dot{C}}{C} \right) \xi^0 + 2 \, \Phi_{,\psi} = 0, \nonumber \\& & \xi^0_{,t} - \xi^1_{,x} + \xi^2_{,y} + \xi^3_{,z} + \left( -\frac{\dot{A}}{A} + \frac{\dot{B}}{B} + \frac{\dot{C}}{C} \right) \xi^0 + 2 \, \Phi_{,\psi} = 0, \label{ngs-cond-3}
\\ & & \xi^0_{,t} + \xi^1_{,x} - \xi^2_{,y} + \xi^3_{,z} + \left( \frac{\dot{A}}{A} - \frac{\dot{B}}{B} + \frac{\dot{C}}{C} \right) \xi^0 + 2 \, \Phi_{,\psi} = 0,  \nonumber \\& & \xi^0_{,t} + \xi^1_{,x} + \xi^2_{,y} - \xi^3_{,z} + \left( \frac{\dot{A}}{A} + \frac{\dot{B}}{B} - \frac{\dot{C}}{C} \right) \xi^0 + 2\, \Phi_{,\psi} = 0, \nonumber \\& &  F_{,t} \xi^0 + F_{,x} \xi^1 + F_{,y} \xi^2 + F_{,z} \xi^3 + \Phi F_{,\psi} +  F \left[ \xi^0_{,t} + \xi^1_{,x} + \xi^2_{,y} + \xi^3_{,z} + \left( \frac{\dot{A}}{A} + \frac{\dot{B}}{B} + \frac{\dot{C}}{C} \right) \xi^0 \right] \nonumber \\& & \qquad  + \frac{1}{A B C} \left( f^0_{,t}  + f^1_{,x} + f^2_{,y} + f^3_{,z} \right) = 0.  \nonumber
\end{eqnarray}
It is noted here that the set of all Noether symmetries with the gauge functions $f^i$ form a finite dimensional Lie algebra. % denoted by $\mathcal{N}$.

%3
\section{Noether Symmetries and Conservation Laws}
\label{sec:3}

It is easily seen from Equation \eqref{ckv-eq} that for arbitrary forms of metric functions $A(t), B(t)$ and $C(t)$, the background spacetime of Bianchi I metric admit the three KVs, the generators of translations in $x$, $y$ and $z$ directions which implies momentum conservation,
\begin{equation}
  {\bf K}_1 = \partial_x, \qquad {\bf K}_2 = \partial_y, \qquad {\bf K}_3 = \partial_z \, . \label{kvs-b1}
\end{equation}
It is obvious that these KVs are also Noether, and so Lie, symmetries of the Klein--Gordon Equation \eqref{kg-eq-2} in the background of Bianchi I spacetime. Hence, applying the expression \eqref{c-vector} of conservation law, the resulting conserved flow vector components related to the Klein--Gordon Equation \eqref{kg-eq-2} are as follows:
\begin{eqnarray}
& & T^t = - A B C \psi_t \psi_x, \,\, T^x = \frac{A B C}{2} \left(  \psi_t^2 + \frac{1}{A^2} \psi_x^2 -  \frac{1}{B^2} \psi_y^2 - \frac{1}{C^2} \psi_z^2 + 2\,  V_1 (t,y,z,\psi) \right) \, , \nonumber \\& &  T^y = \frac{A C}{B} \psi_x \psi_y, \quad  T^z = \frac{A B}{C}  \psi_y \psi_z \, , \label{cl-K0}
\end{eqnarray}
for ${\bf K}_1 = \partial_x$,
\begin{eqnarray}
& & T^t = - A B C \psi_t \psi_y, \,\, T^x = \frac{B C}{A} \psi_x \psi_y,  \nonumber \\& & T^y = \frac{A B C}{2} \left(  \psi_t^2 - \frac{1}{A^2} \psi_x^2 +  \frac{1}{B^2} \psi_y^2 - \frac{1}{C^2} \psi_z^2 + 2\, V_2 (t,x,z,\psi) \right) \, , \quad \label{cl-K1} \\& & T^z = \frac{A B}{C} \psi_y \psi_z. \nonumber
\end{eqnarray}
for ${\bf K}_2 = \partial_y$,
\begin{eqnarray}
& & T^t = - A B C \psi_t \psi_z, \,\, T^x = \frac{B C}{A} \psi_x \psi_z, \,\, T^y = \frac{A C}{B}  \psi_y \psi_z \nonumber \\& & T^z =  \frac{A B C}{2} \left(  \psi_t^2 - \frac{1}{A^2} \psi_x^2 -  \frac{1}{B^2} \psi_y^2 + \frac{1}{C^2} \psi_z^2 + 2\,  V_3 (t,x,y,\psi) \right) \, . \label{cl-K2}
\end{eqnarray}
for ${\bf K}_3 = \partial_z$.
Here, the conserved flow vector is ${\bf T} = (T^t, T^x,T^y,T^z)$, and $V_1 (t,y,z,\psi)$, $V_2 (t,x,z,\psi)$, and $V_3 (t,x,y,\psi)$ are integration functions. We give a complete solution of Noether symmetry conditions \eqref{ngs-cond-3} for the Bianchi I spacetime in the following.

\vspace{3pt} \noindent {\bf Case (i):} First, let us consider the Klein--Gordon equation that requires that $F \neq 0$. Taking $F = U_0 + U_1 \psi + \frac{1}{2} U_2^2 \psi^2$ for any metric coefficients of Bianchi I spacetime, where $U_0, U_1$ and $U_2$ are constants, the components of the Noether symmetry generator \eqref{vecf} are found from Equations \eqref{ngs-cond-3} as
\begin{eqnarray}
  & & \xi^0 = 0 \, , \,\, \xi^1 = c_1 \, , \,\, \xi^2 = c_2 \, , \,\, \xi^3 = c_3 \, ,  \label{nsc-i-1}  \\ & &  \Phi = \left( c_4 \, e^{a_1 x} + c_5 \, e^{-a_1 x} \right) \left( c_6 \, e^{a_2 y} + c_7 \, e^{-a_2 y} \right) \left( c_8 \, e^{a_3 z} + c_{9} \, e^{-a_3 z} \right) Y(t) \, , \label{nsc-i-2}
\end{eqnarray}
along with the gauge vector components
\begin{eqnarray}
  & & f^0 = - A B C \left( c_4 \, e^{a_1 x} + c_5 \, e^{-a_1 x} \right) \left( c_6 \, e^{a_2 y} + c_7 \, e^{-a_2 y} \right) \left( c_8 \, e^{a_3 z} + c_9 \, e^{-a_3 z} \right) \psi \dot{Y}(t) + F_0( t,x,y,z) \, , \nonumber \\ & & f^1 = a_1 \frac{B C}{A} \left( c_4 \, e^{a_1 x} - c_5 \, e^{-a_1 x} \right) \left( c_6 \, e^{a_2 y} + c_7 \, e^{-a_2 y} \right) \left( c_8 \, e^{a_3 z} + c_9 \, e^{-a_3 z} \right) \psi Y(t) + F_1 (t,x,y,z) \, , \nonumber   \\ & & f^2 = a_2 \frac{A C}{B} \left( c_4 \, e^{a_1 x} + c_5 \, e^{-a_1 x} \right) \left( c_6 \, e^{a_2 y} - c_7 \, e^{-a_2 y} \right) \left( c_8 \, e^{a_3 z} + c_9 \, e^{-a_3 z} \right) \psi Y(t) + F_2 (t,x,y,z) \, , \qquad \label{gauge-i} \\ & & f^3 = A B \left( \frac{a_3 \psi}{C} - \frac{U_1 C}{a_3} \right) \left( c_4 \, e^{a_1 x} + c_5 \, e^{-a_1 x} \right) \left( c_6 \, e^{a_2 y} + c_7 \, e^{-a_2 y} \right) \left( c_8 \, e^{a_3 z} - c_9 \, e^{-a_3 z} \right) Y(t) \nonumber  \\ & & \qquad \quad - \int{ ( F_{0,t} + F_{1,x} + F_{2,y} ) } dz +  F_3 (t,x,y) \, , \nonumber
\end{eqnarray}
where $c_1,...,c_9$ and $a_1, a_2, a_3$ are constant parameters, $F_0,...,F_3$ are integration functions, and $Y(t)$ solves the following second-order ordinary differential equation:
\begin{equation}
  \ddot{Y} + \left( \frac{\dot{A}}{A} + \frac{\dot{B}}{B} + \frac{\dot{C}}{C} \right) \dot{Y} - \left( \frac{a_1^2}{A^2} + \frac{a_2^2}{B^2} + \frac{a_3^2}{C^2} + U_2^2 \right) Y (t) = 0 \, .  \label{ode-case-i}
\end{equation}
Therefore, there are {\it eleven} Noether symmetries, such that
\begin{eqnarray}
  & & {\bf X}_1 = {\bf K}_1 \, , \quad {\bf X}_2 = {\bf K}_2 \, , \quad {\bf X}_3 = {\bf K}_3 \, , \label{ngs-i-1} \\ & & {\bf X}_4 = e^{a_1 x + a_2 y + a_3 z} Y (t) \partial_{\psi} \, , \quad {\bf X}_5 = e^{-a_1 x + a_2 y + a_3 z} Y (t) \partial_{\psi} \, , \quad {\bf X}_6 = e^{a_1 x - a_2 y + a_3 z} Y (t) \partial_{\psi} \, , \label{ngs-i-2} \\ & & {\bf X}_7 = e^{a_1 x + a_2 y - a_3 z} Y (t) \partial_{\psi} \, , \qquad {\bf X}_8 = e^{a_1 x - a_2 y - a_3 z} Y (t) \partial_{\psi} \, , \qquad {\bf X}_{9} = e^{-a_1 x + a_2 y - a_3 z} Y (t) \partial_{\psi} \, , \label{ngs-i-3} \\ & &  {\bf X}_{10} = e^{-a_1 x - a_2 y + a_3 z} Y (t) \partial_{\psi} \, , \qquad {\bf X}_{11} = e^{-a_1 x - a_2 y - a_3 z} Y (t) \partial_{\psi} \, . \label{ngs-i-4}
\end{eqnarray}
with the corresponding non-zero gauge vectors:
\begin{eqnarray}
  & & {\bf f}_{4,5} = e^{ \pm a_1 x + a_2 y + a_3 z } \, A B C \, \psi \left( - \dot{Y} , \, \pm \frac{a_1 Y}{A^2}  , \, \frac{a_2 Y}{B^2} , \, \Big{(} \frac{a_3}{C^2} - \frac{U_1}{a_3 \psi} \Big{)} Y  \right) \, , \nonumber \\ & &  {\bf f}_{6,7} = e^{a_1 x \mp a_2 y \pm a_3 z } \, A B C \, \psi \left( - \dot{Y} , \, \frac{a_1 Y}{A^2} , \, \mp \frac{a_2 Y}{B^2} , \, \pm \Big{(} \frac{a_3}{C^2} - \frac{U_1}{a_3 \psi} \Big{)} Y \right)  \, ,  \label{gauge-v-i} \\ & & {\bf f}_{8,9} = e^{ \pm a_1 x \mp a_2 y - a_3 z } \, A B C \, \psi \left( - \dot{Y} , \, \pm \frac{a_1 Y}{A^2} , \, \mp \frac{a_2 Y}{B^2} , \, - \Big{(} \frac{a_3}{C^2} - \frac{U_1}{a_3 \psi} \Big{)} Y \right)  \, , \nonumber \\ & & {\bf f}_{10,11} = e^{-a_1 x - a_2 y \pm a_3 z } \, A B C \, \psi \left( - \dot{Y} , \, - \frac{a_1 Y}{A^2} , \, - \frac{a_2 Y}{B^2} , \, \pm \Big{(} \frac{a_3}{C^2} - \frac{U_1}{a_3 \psi} \Big{)} Y \right)\, . \nonumber
\end{eqnarray}
The conserved vector fields associated with ${\bf X}_1,...,{\bf X}_{11}$ given in \eqref{ngs-i-1}--\eqref{ngs-i-4} are obtained as 
\begin{eqnarray}
    & & {\bf T}_1 = - \frac{1}{2} A B C \,  W \, {\bf K}_1 - \psi_x {\bf T}_0 \, , \,\,  {\bf T}_2 = - \frac{1}{2} A B C \, W \, {\bf K}_2 - \psi_y {\bf T}_0 \, , \,\, {\bf T}_3 = - \frac{1}{2} A B C \, W \, {\bf K}_3 - \psi_z {\bf T}_0 \, , \nonumber  \\ & & {\bf T}_4 = e^{a_1 x + a_2 y + a_3 z } \, Y \, {\bf T}_0  + {\bf f}_4 \, , \, \, {\bf T}_5 = e^{-a_1 x + a_2 y + a_3 z} \, Y \, {\bf T}_0 + {\bf f}_5 \, , \,\,  {\bf T}_6 = e^{a_1 x - a_2 y + a_3 z } \, Y \, {\bf T}_0 + {\bf f}_6 \, , \nonumber \\ & & {\bf T}_7 = e^{a_1 x + a_2 y - a_3 z} \, Y \, {\bf T}_0 + {\bf f}_7 \, , \, \, {\bf T}_8 = e^{ a_1 x - a_2 y - a_3 z } \, Y \, {\bf T}_0 + {\bf f}_8 \, , \quad  {\bf T}_9 = e^{-a_1 x + a_2 y - a_3 z } \, Y \, {\bf T}_0 + {\bf f}_9  \, , \qquad \label{c-vectors-i} \\ & & {\bf T}_{10} = e^{-a_1 x - a_2 y + a_3 z } \, Y \, {\bf T}_0 + {\bf f}_{10} \, , \quad {\bf T}_{11} = e^{-a_1 x - a_2 y - a_3 z } \, Y \, {\bf T}_0 + {\bf f}_{11} \, ,  \nonumber
\end{eqnarray}
where the gauge vectors ${\bf f}_5,...,{\bf f}_{11}$ are the same as in \eqref{gauge-v-i}, and $W$ and ${\bf T}_0$ are defined by
\begin{eqnarray}
  & & W = -\psi_t^2 + \frac{1}{A^2} \psi_x^2 + \frac{1}{B^2} \psi_y^2 + \frac{1}{C^2} \psi_z^2 - 2 \, F \, , \quad {\bf T}_0 = A B C \, \left( \psi_t, - \frac{1}{A^2} \psi_x, - \frac{1}{B^2} \psi_y, - \frac{1}{C^2} \psi_z \right) \, , \label{W-t0-i}
\end{eqnarray}
where $F = U_0 + U_1 \psi + \frac{1}{2} U_2 \psi^2$.
The conserved vector components \eqref{c-vector} for the integration functions $F_0,...,F_3$ of the gauge functions which have the property of $T^i = f^i$ are
\begin{eqnarray}
 & {\bf  T} = \left( F_0 (t,x,y,z) , F_1 (t,x,y,z) , F_2 (t,x,y,z) , -\int{(F_{0,t} + F_{1,x} + F_{2,y}) dz} + F_3 (t,x,y) \right) \, .
\end{eqnarray}
We note, here, that the above conserved quantities will appear in each of the possible cases. Therefore, we will not mention these quantities again.

Through to the end of this section, following Ref. \cite{jamal1}, we will take into account the {\it power-law} form of metric functions, such that
\begin{equation}
  A(t)= t^L \, , \qquad  B(t)= t^p \, , \qquad C(t) = t^q \, , \label{power-law-metric}
\end{equation}
where $L, p$, and $q$ are constant parameters. For the latter forms of the metric functions, we find that there exists an additional HKV, which is a scaling transformation or a dilation,
\begin{equation}
  {\bf K}_4 =  -t \partial_t + (L-1) x \partial_x + (p-1) y \partial_y + (q-1) z \partial_z \, , \label{HV-i}
\end{equation}
where $\sigma = {\rm const.} = 1$, in addition to the KVs ${\bf K}_1, \, {\bf K}_2 $ and ${\bf K}_3$ in \eqref{kvs-b1}. Furthermore, we will consider some subcases in which we obtain the symmetry generators for the wave ($F = U_0 = const.$, where one can take $F=0$ without loss of generality) and Klein--Gordon equations ($F \neq const.$) of Bianchi I spacetime.

Obviously, other choices for the metric functions will lead to a different solution for the function of $Y(t)$ from Equation \eqref{ode-case-i}. We give some examples of these choices: \linebreak {\bf (i.1)} $A(t) = \sin (L t)$, $B(t)= \cos(p t)$, $C(t)= \cos(q t)$; {\bf (i.2)} $A(t) = \sin (L t)$, \mbox{$B(t)= \sin(p t)$}, $C(t)= \sin(q t)$; {\bf (i.3)} $A(t) = \cos(L t)$, $B(t)= \cos(p t)$, $C(t)= \cos(q t)$; and {\bf (i.4)} \mbox{$A(t) = {\rm sech}(L t)$}, $B(t)= {\rm sech}(p t)$, $C(t)= {\rm sech}(q t)$. In subcase {\bf (i.1)}, if $L=p=1$ and $q=0$, then the line element reduces to the conformally flat Bianchi I spacetime, and point symmetries and potentials for the Klein--Gordon equation with $F= V(t,x,y,z) \psi^2 / 2$ in this metric were studied by Ref. \cite{ptm}. The solution $Y(t)$ of \eqref{ode-case-i} for $L = p$ and $q=0$ is found as follows:
\begin{eqnarray}
  & & Y(t) = \frac{ g_2 (t)^{ \frac{p+ a_1}{2 p} } }{\sin (2 p t)} \Big{[} b_1 g_1 (t)^{\frac{d_1}{2} } \, _2F_1 \left( n_1 - d_1, n_2 - d_1 ; 2 d_1;  g_1 (t)  \right) +  b_2 g_1 (t)^{\frac{d_2}{2}} \, _2F_1 \left( n_1 + d_2, n_2 + d_2 ; 2 d_2;  g_1 (t)  \right) \Big{]} \, , \qquad \label{sol-Y-case-i1}
\end{eqnarray}
where $p \neq 0$, $d_1 = 1- a_2 / p$, $d_2 = 1 + a_2 / p$, $n_1 = ( a_1 + \sqrt{p^2 -a_3^2 - U_2^2})/(2 p)$, $n_2 = ( a_1 - \sqrt{p^2 -a_3^2 - U_2^2})/(2 p)$, and the functions $g_1(t), g_2(t)$ are defined as
\begin{equation}
   g_1(t)= \frac{1}{2} \left[ \cos (2 p t) +1 \right] \, , \quad g_2(t)= \frac{1}{2} \left[ \cos (2 p t) - 1 \right] \, . \label{functions-g1-g2}
\end{equation}
The solution \eqref{sol-Y-case-i1} is a new one that was not mentioned in Ref. \cite{ptm} and generalizes the case 4.4 of this reference, where they were taken as $L=p=1$. For $L=p=1$ and $q=3$, we have the following solution for Equation \eqref{ode-case-i}:
\begin{eqnarray}
  & & Y(t) = \frac{ \sin (2t) g_2 (t)^{\frac{a_1}{2}} g_3 (t)^{ \frac{a_3}{36\, d_2} ( 4 a_3 + 3 \sqrt{25 - 4 U_2^2} ) } }{ [2 g_1 (t^)]^{\frac{3}{4}} \sqrt{ 2 \cos^2 (2t) - 3 \cos (2 t) +1 }} \Big{[} b_1 (\cos t)^{d_1} H_G \left( \frac{3}{4}, \alpha_1, \beta_1, \gamma_1, \delta_1, 1 + a_1, g_1 (t) \right)  \nonumber \\ & & \qquad \qquad \qquad \qquad \qquad \qquad \qquad \qquad \qquad \qquad \quad + b_2 (\cos t)^{-d_1} H_G \left( \frac{3}{4}, \alpha_2, \beta_2, \gamma_2, \delta_2, 1 + a_1, g_1 (t) \right) \Big{]} \, ,
\end{eqnarray}
where $H_G$ is the Heun general function, the functions $g_1 (t), g_2(t)$ are the same ones as \mbox{in \eqref{functions-g1-g2}} with $p=1$, and we define that $g_3 (t) = [ \cos(2 t) -1/2]/2$, $\beta_1 = 1 + (a_1 + d_1)/2 + d_2$, $\beta_2 = 1 + (a_1 - d_1)/2 + d_2$, $\delta_1 = 1 + d_1$, $\delta_2 = 1 - d_1$, and
$$ \alpha_1 = \frac{d_1}{8} ( 3 a_1 + 7) + a_1^2 + a_2^2 + \frac{11}{27} a_3^2 + 2 a_1 + U_2^2 - \frac{7}{3} + \frac{a_3 ( 1 + d_1)}{9 \, d_2} ( 4 a_3 + 3 \sqrt{25 - 4 U_2^2} ) ,$$ $$ \alpha_2 = -\frac{d_1}{8} ( 3 a_1 + 7) + a_1^2 + a_2^2 + \frac{11}{27} a_3^2 + 2 a_1 + U_2^2 - \frac{7}{3} + \frac{a_3 ( 1 - d_1)}{9 \, d_2} ( 4 a_3 + 3 \sqrt{25 - 4 U_2^2} )  ,$$
in which $d_1$ and $d_2$ are defined as
$$d_1 = \frac{1}{6} \sqrt{ 9 + 4 ( 9 a_2^2 + a_3^2)}, \qquad d_2 = \frac{1}{12} \sqrt{ 225 + 16 a_3^2 - 36 U_2^2 + 24 a_3 \sqrt{25 - 4 U_2^2}} \, .$$

One can find other solutions of \eqref{ode-case-i} for different choices of the parameters $L, p$ and $q$. Some solutions of the Equation \eqref{ode-case-i} for the remaining subcases {\bf (i.2)}--{\bf (i.4)} are included in Table \ref{Tab1}.

\begin{table*}
\caption{\label{Tab1} Some solutions of the equation \eqref{ode-case-i} for the metric functions given in subcases {\bf (i.2)}, {\bf (i.3)} and {\bf (i.4)}. Here, it is defined that $k = \sqrt{a_1^2 + a_2^2 + a_3^2}$, $k_1 = \sqrt{1 + a_2^2 + a_3^2}$, $k_2 = \left(\sqrt{k^2 + q^2} \right)/(2 q)$, $k_3 = \left(\sqrt{k^2 + 4} \right)/ 4$, $k_4 = \left( \sqrt{a_2^2 + a_3^2 }\, \right)/q$, $k_5 = \left( \sqrt{a_1^2 + q^2}\, \right) /q$, $k_6 = \left( \sqrt{k^2 - U_2^2} \, \right)/q$, $u_1 = \left( \sqrt{9 q^2 - 4 U_2^2} \right) / (4 q) $, $u_2 = \left( \sqrt{4 - U_2^2} \right) / 2$, $u_3 = \sqrt{25 - 4 U_2^2}$ and $u_4 = \sqrt{26 - 4 U_2^2 + 2 u_3}$. We also defined that $\ell_1(t) =  \cos(2 q t) +1$, $\ell_2 (t) = \cos(2 q t) -1$ and $\ell_{3^{\pm}} (t) = (\cos 2 t)^{-\frac{1}{2} (k_1 \pm 1)} $. The special functions $H_G = HeunG \left( -1,\alpha,\beta,\gamma,\delta, \tau(t) \right)$ and $H_C = HeunC \left( \alpha, \beta, \gamma, \delta,\eta, \tau(t) \right)$ are the Heun General function and the Heun Confluent function, respectively. Further, $_2F_1 \left( n_1, n_2; d; \ell (t) \right)$ represents the hypergeometric function. }
\begin{tabular}{c|cc}
Case & $L \, , p \, , q$ & $Y(t)$  \\
\hline \\
${\bf (i.2)}$ & $L =  p = q $ & $ \frac{\left( \ell_2 (t)/2 \right)^{k_2 + \frac{1}{2}}}{\ell_2 (t)} \Big{[} b_1 \,  _2F_1 \left( n, -n; \frac{1}{2}; \frac{1}{2} \ell_1 (t)  \right) + b_2 \sqrt{\ell_1 (t) } \, _2F_1 \left( -n, n; \frac{3}{2}; \frac{1}{2} \ell_1 (t)  \right) \Big{]}  $   \\ & & where $n = k_2 + \frac{1}{4} + u_1$  \\\\ \hline \\
& $L = 2, p= q = 1$ & $\frac{ \sin (2 t) (\cos^2 t -1)^{k_3} }{\ell_2 (t) \sqrt{\ell_1(t)} } \Big{[} b_1 (\cos t)^{-\frac{a_1}{2}} \, _2F_1 \left( n_1, n_2; 2 \, d_{-}; \frac{1}{2} \ell_1 (t)  \right) $  \\ & & $ \qquad \qquad \qquad \qquad \quad + \,\, b_2 \, (\cos t)^{\frac{a_1}{2}} \, _2F_1 \left( n_3, n_4; 2 \, d_{+}; \frac{1}{2} \ell_1 (t)  \right)  \Big{]}$  \\ & & where $d_{\pm} = \frac{1}{2} ( 1 \pm \frac{a_1}{2}), \quad n_1 = k_3 + d_{-} -u_2, \quad n_2 = k_3 + d_{-} + u_2 $  \\ & & $n_3 = k_3 + d_{+} + u_2, \quad n_4 = k_3 + d_{+} - u_2$  \\\\ \hline \\
${\bf (i.3)} $ & $L = p = q $ & $ \frac{\sqrt{\ell_2 (t)}}{\ell_1(t)} \Big{[} b_1 [ 2\, \ell_2 (t)]^{\frac{1}{2} + k_2 } \, _2F_1 \left( n_1, n_2; 1 + 2 k_2; \frac{1}{2} \ell_1 (t)  \right) $  \\ & & $ \qquad \qquad \quad + \,\, b_2 \, [ 2 \, \ell(t)]^{\frac{1}{2} - k_2} \, _2F_1 \left( n_3, n_4; 1 - 2 k_2; \frac{1}{2} \ell_1 (t)  \right)  \Big{]}$  \\ & & where $n_1 = k_2 + \frac{3}{4} - u_1, \quad n_2 = k_2 + \frac{3}{4} + u_1, \quad n_3 = -k_2 + \frac{3}{4} + u_1$, \\ & & $n_4 = -k_2 + \frac{3}{4} - u_1 $  \\\\ \hline \\ & $L= 1, p = q = 2$ & $(\cos t)^{a_1} \ell_2 (t)^{\frac{1 + u_3 + u_4}{4 u_4}} \Big{[} b_1 \, \ell_{3^-} H_G \left( -1,\alpha_1, \beta_1, \gamma_1, \delta_1, \eta, - \cos 2 t \right)$  \\ & & $\qquad \qquad \qquad \qquad \qquad \quad + \,\, b_2 \, \ell_{3^+} H_G \left( -1,\alpha_2, \beta_2, \gamma_2, \delta_2, \eta, - \cos 2 t \right) \Big{]} $  \\ & & where $\lambda_{1^{\pm}} = 1 + \frac{1}{2} ( a_1 \pm k_1 ), \,\, \delta_1 = 1 + k_1, \quad \delta_2 = 1- k_1, \quad \eta = 1 + a_1 $ \\ & & $\alpha_1 = \frac{1}{2} ( 1 - a_1 \delta_1) + \frac{\sqrt{k_1}}{2 u_4} \left[ k_1 (1- k_1) + \frac{1}{2} (1 + u_3) \right] + \frac{1 + u_3}{4 \, u_4}$, \\ & & $\alpha_2 = \frac{1}{2} ( 1 - a_1 \delta_2) - \frac{\sqrt{k_1}}{2 u_4} \left[ k_1 (1- k_1) - \frac{1}{2} (1 + u_3) \right] + \frac{1 + u_3}{4 \, u_4}$,  \\ & & $\beta_1 = \lambda_{1^{+}} + \frac{1}{4} u_4, \,\, \beta_2 = \lambda_{1^{-}} + \frac{1}{4} u_4, \,\, \gamma_1 = \lambda_{1^+} + \frac{U_2^2 -6}{u_4}, \,\, \gamma_2 = \lambda_{1^{-}} + \frac{U_2^2 -6}{u_4}$  \\\\ \hline \\
${\bf (i.4)} $ & $L = p = q$ & $\cosh^4 (q t) \Big{[} b_1 H_C \left( 0,-\frac{1}{2},2,\frac{k^2}{4 q^2}, \frac{3 q^2 - k^2 - U_2^2}{4 \, q^2}, -\sin^2 t \right) $  \\ &  & $ \qquad \qquad \qquad \qquad \quad + \,\,  b_2 \sinh (q t) H_C \left( 0,\frac{1}{2},2,\frac{k^2}{4 \, q^2}, \frac{3 q^2 - k^2 - U_2^2}{4 q^2}, -\sin^2 t \right) \Big{]}$ \\\\ \hline \\ & $L=0$, $p = q$ & $b_1 H_C\left( 0, -\frac{3}{2}, -\frac{1}{2}, - \frac{k_4^2}{4},  \frac{5 q^2 - 2 a_1^2 -2 U_2^2}{8 \, q^2}, \cosh^2 (q t) \right)$ \\  & & $ \qquad \qquad \qquad + \,\, b_2 \,\cosh^3 (q t) H_C\left( 0, \frac{3}{2}, -\frac{1}{2}, -\frac{k_4^2}{4},  \frac{5 q^2- 2 a_1^2 - 2 U_2^2}{8 \, q^2}, \cosh^2 (q t) \right) $ \\\\ \hline
\end{tabular}
\end{table*}

\noindent {\bf Case (ii)}: If $F = U_0 + U_1 \psi + \frac{1}{2} U_2^2 \psi^2$, it is found for the power-law form of Bianchi I spacetime that {\it twelve} Noether symmetries appear: 
\begin{eqnarray}
    & & {\bf X}^1_1 = {\bf K}_1  , \qquad \qquad {\bf X}^1_2 = {\bf K}_2, \, \qquad {\bf X}^1_3 = {\bf K}_3, \label{ngs-i1-1} \\& &   {\bf X}^1_4 = {\bf K}_4 + \psi \partial_{\psi} \, , \quad {\bf X}^1_5 = {\bf X}_4 \, , \qquad  {\bf X}^1_6 = {\bf X}_5 \, , \quad  {\bf X}^1_7 = {\bf X}_6 \, , \label{ngs-i1-2} \\ & & {\bf X}^1_8 = {\bf X}_7 \, , \qquad \qquad {\bf X}^1_9 = {\bf X}_8 \, , \qquad {\bf X}^1_{10} ={\bf X}_9 \, , \quad {\bf X}^1_{11} = {\bf X}_{10} \, , \quad {\bf X}^1_{12} = {\bf X}_{11}\, , \label{ngs-i1-3}
\end{eqnarray}
with the corresponding non-zero gauge vectors:
\begin{eqnarray}
  & & {\bf f}^1_{5,6} = e^{ \pm a_1 x + a_2 y + a_3 z } f_0 \left( - \dot{Y} , \, \pm a_1 t^{-2 L} Y , \, a_2 t^{-2 p} Y , \, \left( a_3 t^{-2 q} - \frac{U_1}{a_3 \psi} \right) Y \right) \, , \nonumber \\  & &  {\bf f}^1_{7,8} = e^{a_1 x \mp a_2 y \pm a_3 z } \, f_0 \left( - \dot{Y} , \, a_1 t^{-2 L} Y , \, \mp a_2 t^{-2 p} Y , \, \pm \left( a_3 t^{-2 q} - \frac{U_1}{a_3 \psi} \right) Y \right)  \, ,  \label{gauge-v-i1} \\ & & {\bf f}^1_{9,10} = e^{ \pm a_1 x \mp a_2 y - a_3 z } \, f_0 \left( - \dot{Y} , \, \pm a_1 t^{-2 L} Y , \, \mp a_2 t^{-2 p} Y , \, - \left( a_3 t^{-2 q} - \frac{U_1}{a_3 \psi} \right) Y \right)  \, , \nonumber \\ & & {\bf f}^1_{11,12} = e^{-a_1 x - a_2 y \pm a_3 z } \, f_0 \left( - \dot{Y} , \, - a_1 t^{-2 L} Y , \, - a_2 t^{-2 p} Y , \, \pm \left( a_3 t^{-2 q} - \frac{U_1}{a_3 \psi} \right) Y \right)\, , \nonumber
\end{eqnarray}
where ${f}_0 = t^{L + p + q} \psi$ and ${\bf X}_4,...,{\bf X}_{11}$ are the same as in \eqref{ngs-i-2}--\eqref{ngs-i-4}. Meanwhile, in this subcase, the second-order ordinary differential Equation \eqref{ode-case-i} for $Y(t)$ becomes
\begin{equation}
  \ddot{Y} + \frac{(L+ p+ q)}{t} \dot{Y} - \left( \frac{a_1^2}{t^{2 L}} + \frac{a_2^2}{t^{2 p}} + \frac{a_3^2}{t^{2 q}} + U_2^2 \right) Y (t) = 0 \, .  \label{ode-case-i1-1}
\end{equation}
The conserved flow vectors associated with the KVs ${\bf X}^1_1, {\bf X}^1_2, {\bf X}^1_3$ and non-Killing Noether symmetries ${\bf X}^1_4,...,{\bf X}^1_{12}$ given in \eqref{ngs-i1-2} and \eqref{ngs-i1-3} are obtained as 
\begin{eqnarray}
  & & {\bf T}^1_1 = - \frac{1}{2} t^{L + p + q} \, W \, {\bf K}_1 - \psi_x {\bf T}_0  \, , \quad {\bf T}^1_2 = - \frac{1}{2} t^{L + p + q} \, W \, {\bf K}_2 - \psi_y {\bf T}_0 \, , \label{c-vectors-i1-t1}  \\ & & {\bf T}^1_3 = - \frac{1}{2} t^{L + p + q} \, W \, {\bf K}_3 - \psi_z {\bf T}_0 \, , \quad {\bf T}^1_4 = - \frac{1}{2} t^{L + p + q} \, W \, {\bf K}_4 + ( \psi - Q_0 ) {\bf T}_0 \, , \label{c-vectors-i1-t2}\\ & & {\bf T}^1_5 = e^{a_1 x + a_2 y + a_3 z } \, Y \, {\bf T}_0  + {\bf f}^1_5 \, , \, \, {\bf T}^1_6 = e^{-a_1 x + a_2 y + a_3 z} \, Y \, {\bf T}_0 + {\bf f}^1_6 \, , \,\,  {\bf T}^1_7 = e^{a_1 x - a_2 y + a_3 z } \, Y \, {\bf T}_0 + {\bf f}^1_7 \, , \label{c-vectors-i1-t3} \\ & & {\bf T}^1_8 = e^{a_1 x + a_2 y - a_3 z} \, Y \, {\bf T}_0 + {\bf f}^1_8 \, , \, \, {\bf T}^1_9 = e^{ a_1 x - a_2 y - a_3 z } \, Y \, {\bf T}_0 + {\bf f}^1_9 \, , \quad  {\bf T}^1_{10} = e^{-a_1 x + a_2 y - a_3 z } \, Y \, {\bf T}_0 + {\bf f}^1_{10}  \, , \qquad \label{c-vectors-i1-t4} \\ & & {\bf T}^1_{11} = e^{-a_1 x - a_2 y + a_3 z } \, Y \, {\bf T}_0 + {\bf f}^1_{11} \, , \,\, {\bf T}^1_{12} = e^{-(a_1 x + a_2 y + a_3 z )} \, Y \, {\bf T}_0 + {\bf f}^1_{12} \, ,  \label{c-vectors-i1-t5}
\end{eqnarray}
where $W, {\bf T}_0$ are given in Equation \eqref{W-t0-i} by taking $A = t^L, B = t^p, C = t^q$, $Q_0$ is defined as
\begin{eqnarray}
  Q_0 = - t \psi_t + (L-1) x \psi_x + (p-1) y \psi_y + (q-1) z \psi_z \, , \label{Q-i1}
\end{eqnarray}
and the gauge vectors ${\bf f}^1_5,...,{\bf f}^1_{12}$ have the same forms as in \eqref{gauge-v-i1}.
Furthermore, this subcase yields a thirteenth Noether symmetry, in addition to the twelve obtained above, which is ${\bf X} = y \partial_x - x \partial_y$ if $L = p$, and $q$ is an arbitrary constant; ${\bf X} = z \partial_x - x \partial_z$ if $L = q$, and $p$ is an arbitrary constant, and finally ${\bf X} = z \partial_y - y \partial_z$ if $p = q$, and $L$ is an arbitrary constant.

In this case, let us examine some solutions of Equation \eqref{ode-case-i1-1} for specific values of the constant powers $L, p,$ and $q$ of the metric functions. If we assume $L=2,\, p=q=0$, the solution of the Equation \eqref{ode-case-i1-1} is expressed in terms of double confluent Heun functions:
\begin{equation}
  Y(t) = \frac{1}{\sqrt{t}} \left[ b_1 H_D  + b_2 H_D \int{ \frac{dt}{t\, H_D^2} } \right] \, ,
\end{equation}
where $b_1, b_2$ are constant parameters, and $H_D = HeunD \left( \alpha,\beta,\gamma,\delta, \frac{t^2+1}{t^2-1} \right)$ with $\alpha =0$, $\beta = -\frac{1}{4} - (a_1^2 + a_2^2 + a_3^2) - U_2^2$, $\gamma = 2 ( a_1^2 - a_2^2 - a_3^2 - U_2^2)$, and $\delta = \frac{1}{4} - (a_1^2 + a_2^2 + a_3^2) - U_2^2$. For $L=-3,\, p=q=0$, we have the solution
\begin{equation}
  Y(t) = t^4 e^{ \frac{a_1 t^4}{4}} \left[ b_1 H_B  + b_2 H_B \int{ \frac{ e^{ \frac{a_1 t^4}{2}} dt}{t^5 \, H_B^2} } \right] \, ,
\end{equation}
where $H_B$ is the Heun biconfluent function such that $H_B = HeunB \left( \alpha,\beta,\gamma,\delta, \sqrt{\frac{a_1}{2}} t^2 \right)$ with $\alpha = 2$, $\beta =0$, $\gamma = 0$, and $\delta = \frac{1}{\sqrt{2 a_1}} (a_3^2 + U_2^2)$. Further analysis of Equation \eqref{ode-case-i1-1} with respect to values of $L, p$, and $q$ reveals non-trivial solutions in terms of some familiar special functions. Table \ref{Tab2} contains some of these solutions corresponding to specific values for $L, p$, and $q$.

\begin{table*}
\caption{\label{Tab2} Some solutions of the equation \eqref{ode-case-i1-1} for specific values of $L, p$, and $q$. Here, it is defined that $\epsilon_1 = \sqrt{a_2^2 + a_3^2 + U_2^2}, \epsilon_2 = \sqrt{a_3^2 + U_2^2}, \eta_1 = \sqrt{1 + 4 a_1^2} $ and $\eta_2 = \sqrt{1 + 16 a_1^2}$. Further, the special functions $K_M = KummerM (\mu,\nu,z)$ and $K_U = KummerU (\mu,\nu,z)$ are the Kummer functions, $W_M = WhittakerM (\mu,\nu,z)$ and $W_W = WhittakerW (\mu,\nu,z)$ are the Whittaker functions, respectively. }
\begin{tabular}{c|cc}
Case & $L \, , p \, , q$ & $Y(t)$ \\
\hline \\
$1 $ & $L = \frac{1}{2}, p = q = 0$ & $ \sqrt{t}\, e^{- \epsilon_1 t} \left[ b_1 K_M (\mu, \nu, 2 \epsilon_1 t ) + b_2 K_U (\mu, \nu, 2 \epsilon_1 t ) \right] $  \\ & & where $\mu = \frac{a_1^2}{2 \, \epsilon_1} + \frac{3}{4}, \,\, \nu = \frac{3}{2}$.  \\\\ \hline \\
$2 $ & $L=-1, p = q = 0$ & $ t^2 e^{-\frac{a_1 t^2}{2}} \left[ b_1 K_M (\mu, \nu, a_1 t^2) + b_2 K_U (\mu, \nu, a_1 t^2) \right] $  \\ & & where $\mu = 1 + \frac{\epsilon_1^2}{4 a_1}, \,\,\nu = 2$.  \\\\ \hline \\
$3 $ & $L = 1, p= -1, q= 0$ & $\frac{1}{\sqrt{t}} \left[ b_1 W_M (\mu, \nu, a_2 t^2) + b_2 W_W (\mu, \nu, a_2 t^2) \right] $ \\ & & where $\mu = -\frac{\epsilon_2^2}{4 a_2}, \,\, \nu = \frac{1}{4} \eta_1 $.  \\\\ \hline \\
$4 $ & $L = 1, p = \frac{1}{2}, q= 0 $ & $ t^{-\frac{3}{4}} \left[ b_1 W_M (\mu, \nu, 2\, \epsilon_2 t) + b_2 W_W (\mu, \nu, 2 \, \epsilon_2 t) \right] $  \\ & & where $\mu = -\frac{a_2^2}{2\, \epsilon_2 } \, , \,\, \nu = \frac{1}{4}\eta_2 $.  \\\\ \hline \\
$5 $ & $L = 1, p = -1, q = -3$ & $t^2 e^{-\frac{a_3 t^4}{4}} \Big{[} b_1 t^{\alpha} H_B (\alpha, \beta, \gamma, \delta, \sqrt{\frac{a_3}{2}} t^2)  + b_2 t^{-\alpha} H_B (-\alpha, \beta, \gamma, \delta, \sqrt{\frac{a_3}{2}} t^2) \Big{]}$  \\ & & where $\alpha = \sqrt{4 + a_1^2}, \,\, \beta = 0, \,\, \gamma = -\frac{a_2^2}{2 a_3}, \,\, \delta = \frac{U_2^2}{\sqrt{2 a_3}}$ \\\\ \hline
$6 $ & $L = 3, p = 2, q = -1$ & $t^{-\frac{3}{2}}  \Big{[} b_1 \, e^{-\frac{(a_3 t^4 - a_1)}{2 \, t^2}} H_D (\alpha, \beta, \gamma, \delta, \frac{\alpha t^2 - 4 a_1}{\alpha t^2 + 4 a_1} ) $ \\ &  & $\qquad \qquad + \,\, b_2 \, e^{\frac{(a_3 t^4 - a_1)}{2 \, t^2}} H_D (-\alpha, \beta, \gamma, \delta, \frac{\alpha t^2 - 4 a_1}{\alpha t^2 + 4 a_1} ) \Big{]}$  \\ & & where $\alpha = 4 \sqrt{a_1 a_3}, \quad \beta = \frac{\alpha a_2^2}{4 a_1} + \frac{\alpha^2}{8}- \frac{9}{4} + U_2^2 \sqrt{\frac{a_1}{a_3}},$ \\ & & $\gamma = -\frac{\alpha a_2^2}{2 a_1} + 2 \, U_2^2 \sqrt{\frac{a_1}{a_3}}, \quad \delta = \frac{\alpha a_2^2}{4 a_1} -\frac{\alpha^2}{8} + \frac{9}{4} + U_2^2 \sqrt{\frac{a_1}{a_3}}$. \\\\ \hline
$7 $ & $L = -1, p = 3, q = 1$ & $ \frac{1}{t} \Big{[} b_1 \, e^{-\frac{(a_1 t^4 - a_2)}{2 \, t^2}} H_D (\alpha, \beta, \gamma, \delta, \frac{\alpha t^2 - 4 a_2}{\alpha t^2 + 4 a_2} ) $  \\ &  & $\qquad \qquad + \,\, b_2 \, e^{\frac{(a_1 t^4 - a_2)}{2 \, t^2}} H_D (-\alpha, \beta, \gamma, \delta, \frac{\alpha t^2 - 4 a_2}{\alpha t^2 + 4 a_2} ) \Big{]}$  \\ & & $\alpha = 4 \sqrt{a_1 a_2}, \quad \beta = \frac{\alpha^2}{8}  - a_3^2 - 1 + U_2^2 \sqrt{\frac{a_2}{a_1}}$  \\ & & $\gamma =  2 \, U_2^2 \sqrt{\frac{a_2}{a_1}}, \quad \delta = - \frac{\alpha^2}{8} + a_3^2 + 1 + U_2^2 \sqrt{\frac{a_2}{a_1}}$.  \\\\ \hline
\end{tabular}
\end{table*}

\noindent {\bf Subcase (ii.1)}. If we take $F = U_0 + U_1 \psi$ in the Klein--Gordon Equation \eqref{kg-eq-2} for the power law form of Bianchi I spacetime \eqref{power-law-metric} with $L \neq p \neq q$, where $U_0, U_1$ are constants, it is found that there are {\it fifteen} Noether symmetries, as follows:
\begin{eqnarray}
  & & {\bf X}^5_1 = {\bf K}_1 \, , \qquad {\bf X}^5_2 = {\bf K}_2 \, , \qquad {\bf X}^5_3 = {\bf K}_3 \, , \label{ngs-i2-1} \\ & & {\bf X}^5_{4,5} = m\, e^{-a_1 x + a_2 y \pm a_3 z} Y (t) \partial_{\psi} \, , \qquad {\bf X}^5_{6,7} = m \, e^{-a_1 x - a_2 y \pm a_3 z} Y (t) \partial_{\psi} \, , \\ & & {\bf X}^5_{8,9} = ( e^{a_1 x} + m \, e^{-a_1 x} ) \, e^{\, a_2 y \pm a_3 z} Y (t) \partial_{\psi} \, , \quad {\bf X}^5_{10,11} = ( e^{a_1 x} + m \, e^{-a_1 x} ) e^{- a_2 y \pm a_3 z} Y (t) \partial_{\psi} \, , \label{ngs-i2-2} \\ & & {\bf X}^5_{12,13} = {\bf K}_4 + {\bf X}^5_{4,5} + \left( \psi - \frac{3 V_1 t^2}{2(1+ L + p+ q)} - \frac{3 V_1 t^{1-L-p-q}}{2 m (1+L+p+q)} \right)  \partial_{\psi} \, , \label{ngs-i2-4} \\ & & {\bf X}^5_{14,15} = {\bf K}_4 + {\bf X}^5_{6,7} + \left( \psi - \frac{3 V_1 t^2}{2(1+ L + p+ q)} \right)  \partial_{\psi} \, , \label{ngs-i2-5}
\end{eqnarray}
and the corresponding non-zero gauge vectors are
\begin{eqnarray}
   & & {\bf f}^5_{4,5} = m \, e^{ - a_1 x + a_2 y \pm a_3 z } \, f_0 \left( - \dot{Y} ,  - a_1 t^{-2 L} \, Y , \, a_2 t^{-2 p} \, Y , \, \pm \Big{(} a_3 t^{-2 q} - \frac{U_1}{a_3 \psi} \Big{)} Y  \right) \, , \nonumber \\ & & {\bf f}^5_{6,7} = m \, e^{- a_1 x - a_2 y \pm a_3 z } \, f_0 \left( - \dot{Y} , \, - a_1 t^{-2 L} \, Y , \, - a_2 t^{-2 p} \, Y , \, \pm \Big{(} a_3 t^{-2 q} - \frac{U_1}{ a_3 \psi} \Big{)} Y \right)  \, ,  \nonumber \\ & & {\bf f}^5_{8,9} = ( e^{a_1 x} + m \, e^{-a_1 x} ) \,  e^{ \, a_2 y \pm a_3 z } \, f_0 \left( - \dot{Y} , \,  \frac{a_1 ( e^{a_1 x} - m \, e^{-a_1 x} ) }{ ( e^{a_1 x} + m \, e^{-a_1 x} ) }  Y, \, a_2 t^{-2 p} \, Y , \, \pm \Big{(} a_3 t^{-2 q} - \frac{U_1}{a_3 \psi} \Big{)} Y \right) , \nonumber \\ & & {\bf f}^5_{10,11} = ( e^{a_1 x} + m \, e^{-a_1 x} ) \,  e^{ -a_2 y \pm a_3 z } \, f_0 \left( - \dot{Y} , \,  \frac{a_1 ( e^{a_1 x} - m \, e^{-a_1 x} ) }{ ( e^{a_1 x} + m \, e^{-a_1 x} ) }  Y, \, - a_2 t^{-2 p} \, Y , \, \pm \Big{(} a_3 t^{-2 q} - \frac{U_1}{a_3 \psi} \Big{)} Y \right) ,  \nonumber \\  & & {\bf f}^5_{12,13} = {\bf f}^5_{4,5} + \frac{3 U_1 \psi}{ 1 + L + p +q} \left[ t^{1 + L + p + q} +  \frac{(1 - L - p - q)}{2 m} \right] \partial_t   \label{gauge-v-i2} \\ & & \qquad \qquad \quad  + \left[ \frac{3 U_1 t}{2 m ( 1 + L+ p + q)} + t^{L + p + q} \left( 4 U_0 + \frac{3 U_1^2 t^2}{2 ( 1+ L + p + q)} \right) \right] z \partial_z \, , \nonumber \\ & & {\bf f}^5_{14,15} = {\bf f}^5_{6,7} + \frac{3 U_1 \psi \,  t^{1 + L + p + q} }{ 1 + L + p +q}  \partial_t + \, t^{L + p + q} \left[ 4 U_0 + \frac{3 U_1^2 t^2}{2 ( 1+ L + p + q)} \right] z \partial_z \, , \nonumber
\end{eqnarray}
where $f_0 = t^{L + p +q} \psi$, $m$ is a non-zero constant parameter, and ${\bf K}_4$ is the HKV given in \eqref{HV-i}. Thus, one can write the conserved flow vectors for the Noether symmetries ${\bf X}^5_{4},...,{\bf X}^5_{15}$ as
\begin{eqnarray}
  & & {\bf T}^5_{4,5} = m\,Y \,  e^{-a_1 x + a_2 y \pm a_3 z} \, {\bf T}_0 + {\bf f}^5_{4,5} \, ,  \quad {\bf T}^5_{6,7} = m \, Y \, e^{-a_1 x - a_2 y \pm a_3 z} \, {\bf T}_0 + {\bf f}^5_{6,7} \nonumber \\ & & {\bf T}^5_{8,9} = Y\, ( e^{a_1 x} + m \, e^{-a_1 x} ) \, e^{\, a_2 y \pm a_3 z} \, {\bf T}_0 + {\bf f}^5_{8,9} \, , \,\, {\bf T}^5_{10,11} = Y \, ( e^{a_1 x} + m \, e^{-a_1 x} ) \, e^{\, -a_2 y \pm a_3 z} \, {\bf T}_0 + {\bf f}^5_{10,11} \nonumber \\ & & {\bf T}^5_{12,13} =  \left[ m \, Y \, e^{-a_1 x - a_2 y \pm a_3 z} - \frac{3 U_1 t^2}{2 ( 1 + L + p + q)} - \frac{3 U_1 t^{1 - L - p -q}}{2 m (1 + L + p + q)} + \psi \right] {\bf T}_0 \label{c-vectors-i2} \\ & & \qquad \qquad  - \frac{1}{2} t^{L + p + q} \, W \, {\bf K}_4 + {\bf f}^5_{12,13}  \nonumber \\ & & {\bf T}^5_{14,15} =  \left[ m \, Y \, e^{-a_1 x + a_2 y \pm a_3 z} - \frac{3 U_1 t^2}{2 ( 1 + L + p + q)} + \psi \right] {\bf T}_0 - \frac{1}{2} t^{L + p + q} \, W \, {\bf K}_4  + {\bf f}^5_{14,15} \, , \nonumber
\end{eqnarray}
where $W = -\psi_t^2 + t^{-2 L} \psi_x^2 + t^{-2 p} \psi_y^2 + t^{-2 q} \psi_z^2 - 2 ( U_0 + U_1 \psi)$.

\noindent {\bf  Subcase (ii.2)}. For $F = 0$, which requires $U_0 = U_1= U_2 =0$,  Equation \eqref{kg-eq-2} reduces to the wave equation for Bianchi I spacetime. If we take $L = p = q$ in \eqref{power-law-metric}, the Bianchi I spacetime yields the flat FLRW spacetime. For the latter assumption of metric coefficients, and assuming $F = U_0$, we find {\it twenty six} Noether symmetries, which are the KVs ${\bf X}^1_1, {\bf X}^1_2, {\bf X}^1_3$ given in \eqref{ngs-i1-1}, and
\begin{eqnarray}
  & & {\bf X}^2_4 = {\bf K}_4 + \psi \partial_{\psi} \, , \quad {\bf X}^2_5 = {\bf K}_5 \, , \quad {\bf X}^2_6 = {\bf K}_6 \, , \quad {\bf X}^2_7 = {\bf K}_7 \, , \label{ngs-i1-1-4567} \\ & &  {\bf X}^2_8 = {\bf K}^1_8 +  2 (q-1) x \psi \partial_{\psi} \, , \quad  {\bf X}^2_9 = {\bf K}^1_9 + 2 (q-1) y \psi \partial_{\psi} \, , \quad {\bf X}^2_{10} = {\bf K}^1_{10} + 2 (q-1) z \psi \partial_{\psi} \, ,  \label{ngs-i1-1-8-9-10}
  \\ & & {\bf X}^2_{11} = M \, e^{ b_1 x + b_2 y} \sin (\ell z) \partial_{\psi} \, , \quad {\bf X}^2_{12} = M \, e^{ -b_1 x + b_2 y} \sin (\ell z) \partial_{\psi} \, ,  \label{ngs-i1-1-11-12}  \\ & & {\bf X}^2_{13} = M \, e^{ b_1 x - b_2 y} \sin (\ell z) \partial_{\psi} \, , \quad {\bf X}^2_{14} = M \, e^{-( b_1 x + b_2 y )} \sin (\ell z) \partial_{\psi} \, , \label{ngs-i1-1-13-14}
  \\ & & {\bf X}^2_{15} = M \, e^{ b_1 x + b_2 y} \cos (\ell z) \partial_{\psi} \, , \quad {\bf X}^2_{16} = M \, e^{ - b_1 x + b_2 y} \cos (\ell z) \partial_{\psi} \, , \label{ngs-i1-1-15-16} \\ & & {\bf X}^2_{17} = M \, e^{ b_1 x - b_2 y} \cos (\ell z) \partial_{\psi} \, , \quad {\bf X}^2_{18} = M \, e^{-( b_1 x + b_2 y )} \cos (\ell z) \partial_{\psi} \, , \label{ngs-i1-1-17-18}
  \\ & & {\bf X}^2_{19} = N \, e^{ b_1 x + b_2 y} \sin (\ell z) \partial_{\psi} \, , \quad \, {\bf X}^2_{20} = N \, e^{ - b_1 x + b_2 y} \sin (\ell z) \partial_{\psi} \, , \label{ngs-i1-1-19-20}  \\ & & {\bf X}^2_{21} = N \, e^{ b_1 x - b_2 y} \sin (\ell z) \partial_{\psi} \, , \quad \, {\bf X}^2_{22} = N \, e^{-( b_1 x + b_2 y )} \sin (\ell z) \partial_{\psi} \, , \label{ngs-i1-1-21-22}
  \\ & & {\bf X}^2_{23} = N \, e^{ b_1 x + b_2 y} \cos (\ell z) \partial_{\psi} \, , \quad \, {\bf X}^2_{24} = N \, e^{ - b_1 x + b_2 y} \cos (\ell z) \partial_{\psi} \, , \label{ngs-i1-1-23-24}  \\ & & {\bf X}^2_{25} = N \, e^{ b_1 x - b_2 y} \cos (\ell z) \partial_{\psi} \, , \quad \, {\bf X}^2_{26} = N \, e^{-( b_1 x + b_2 y )} \cos (\ell z) \partial_{\psi} \, , \label{ngs-i1-1-25-26}
\end{eqnarray}
in which the vector fields ${\bf K}_5, {\bf K}_6, {\bf K}_7, {\bf K}^1_8, {\bf K}^1_9$ and ${\bf K}^1_{10}$ have the form
\begin{eqnarray}
  & & {\bf K}_5 = y \partial_x - x \partial_y \, , \qquad {\bf K}_6 = z \partial_x - x \partial_z \, , \qquad {\bf K}_7 =  z \partial_y - y \partial_z \, ,  \label{kvs-i1-1-567} \\ & & {\bf K}^1_8 = 2 (q-1) x \left[ - t \partial_t + (q-1) ( y \partial_y + z \partial_z ) \right] + \left[ (q-1)^2 (x^2 - y^2 - z^2 ) + t^{2(1-q)} \right] \partial_x \, , \qquad  \label{sckv-i1-1-8} \\ & & {\bf K}^1_9 = 2 (q-1) y \left[ - t \partial_t + (q-1) ( x \partial_x + z \partial_z ) \right] + \left[ (q-1)^2 (-x^2 + y^2 - z^2 ) + t^{2(1-q)} \right] \partial_y \, , \qquad \label{sckv-i1-1-9} \\ & & {\bf K}^1_{10} = 2 (q-1) z \left[ - t \partial_t + (q-1) ( x \partial_x + y \partial_y ) \right] + \left[ (q-1)^2 (-x^2 - y^2 + z^2 ) + t^{2(1-q)} \right] \partial_z \, , \qquad \label{sckv-i1-1-10}
\end{eqnarray}
and $M = M (t)$ and $N = N (t)$ are defined by
\begin{eqnarray}
  & & M (t) = \frac{(q+1)}{b_3} t^{-\frac{(q+1)}{2} } J_{\nu_1} (\tau ) + t^{ \frac{(1- 3 q)}{2} } J_{\nu_2} (\tau ) \, , \,\, N (t) = \frac{(q+1)}{b_3} t^{-\frac{(q+1)}{2} } Y_{\nu_1} (\tau ) + t^{ \frac{(1-3q)}{2} } Y_{\nu_2} (\tau ) \, , \qquad \label{defns-i1-1-12}
\end{eqnarray}
where $b_1, b_2$ and $b_3 (\neq 0)$ are constant parameters, $q \neq 0, 1, 1/2$, $\nu_1 = -(q+1)/(2 (q-1)),$ $\nu_2 = (q-3)/(2(q-1)), \ell = \sqrt{ b_1^2 + b_2^2 + b_3^2}$, $\tau = (b_3/(q-1)) t^{1-q}$,  $\{ J_{\nu_1} (\tau), J_{\nu_2} (\tau) \}$ and $\{ Y_{\nu_1} (\tau), Y_{\nu_2} (\tau ) \}$ are  first and second kind Bessel functions, respectively.
We note here that the vector fields ${\bf K}_5, {\bf K}_6, {\bf K}_7$ in \eqref{kvs-i1-1-567} are the KVs, and the other ones ${\bf K}^1_8, {\bf K}^1_9$, and ${\bf K}^1_{10}$ in \eqref{sckv-i1-1-8}--\eqref{sckv-i1-1-10} are the SCKVs with the conformal factors $\sigma = 2 (q -1) x, \sigma= 2 (q-1) y$, and $\sigma= 2 (q-1) z$, respectively.
Furthermore, the gauge vectors for ${\bf X}^2_{8},...,{\bf X}^2_{26}$ are
\begin{eqnarray}
  & & {\bf f}^2_{8} = (q-1) t^q \psi^2 \, \partial_x \, , \quad  {\bf f}^2_{9} = (q-1) t^q \psi^2 \, \partial_y \, , \quad {\bf f}^2_{10} = (q-1) t^q \psi^2 \, \partial_z \, , \nonumber \\
  & & {\bf f}^2_{11,12} = \psi \, e^{ \pm b_1 x + b_2 y } \sin (\ell z) \left(  b_3 \, t^{\frac{q+1}{2}} \, J_{\nu_1} (\tau) \, , \, t^q \, M  \Big{(} \pm b_1 \, , \,  b_2 \, , \, \ell \cot(\ell z) \Big{)} \right) \, , \nonumber\\
  & & {\bf f}^2_{13,14} = \psi \, e^{\pm b_1 x - b_2 y } \sin (\ell z) \left(  b_3 \,  t^{\frac{q+1}{2}} J_{\nu_1} (\tau) \, , \, t^q \, M   \Big{(} \pm b_1 \, , \, - b_2 \, , \, \ell \cot(\ell z) \Big{)} \right) \, ,  \nonumber\\
  & & {\bf f}^2_{15,16} =  \psi \, e^{\pm b_1 x + b_2 y } \cos (\ell z) \left(  b_3 \,  t^{\frac{q+1}{2}} J_{\nu_1} (\tau) \, , \, t^q \, M \Big{(} \pm b_1 \, , \,  b_2 \, , \, - \ell \tan (\ell z) \Big{)} \right)  \, , \nonumber\\
  & & {\bf f}^2_{17,18} = \psi \, e^{\pm b_1 x - b_2 y } \cos (\ell z) \left( b_3 \,  t^{\frac{q+1}{2}} J_{\nu_1} (\tau) \, , \, t^q \, M \Big{(} \pm b_1 \, , \, - b_2 \, , \, -\ell \tan(\ell z) \Big{)} \right)  \, ,  \label{gauge-v-ii-2}\\
  & & {\bf f}^2_{19,20} = \psi \, e^{\pm b_1 x + b_2 y } \sin (\ell z) \left( b_3 \,  t^{\frac{q+1}{2}} Y_{\nu_1} (\tau) \, , \, t^q \, N \Big{(} \pm b_1 \, , \, b_2 \, , \,  \ell \cot(\ell z) \Big{)} \right) \, , \nonumber\\
  & & {\bf f}^2_{21,22} = \psi \, e^{ \pm b_1 x - b_2 y } \sin (\ell z) \left( b_3 \,  t^{\frac{q+1}{2}} Y_{\nu_1} (\tau) \, , \, t^q \, N \Big{(} \pm b_1 \, , \, - b_2 \, , \, \ell \cot(\ell z) \Big{)} \right) \, ,  \nonumber \\
  & & {\bf f}^2_{23,24} = \psi \, e^{\pm b_1 x + b_2 y } \cos (\ell z) \left( b_3 \,  t^{\frac{q+1}{2}} Y_{\nu_1} (\tau) \, , \, t^q \, N \Big{(} \pm b_1 \, , \, b_2 \, , \, -\ell \tan (\ell z) \Big{)} \right)  \, , \nonumber \\
 & & {\bf f}^2_{25,26} = \psi \, e^{\pm b_1 x - b_2 y } \cos (\ell z) \left( b_3 \,  t^{\frac{q+1}{2}} Y_{\nu_1} (\tau) \, , \, t^q \, N \Big{(} \pm b_1 \, , \, - b_2 \, , \, -\ell \tan(\ell z) \Big{)} \right)  \, .  \nonumber
\end{eqnarray}

Then, in addition to the conserved flow vectors for ${\bf X}_1, {\bf X}_2, {\bf X}_3$, and ${\bf X}^2_4$ obtained in \eqref{c-vectors-i1-t1} and \eqref{c-vectors-i1-t2}, the conserved vector fields for the remaining Noether symmetries of this subcase become
\begin{eqnarray}
    & & {\bf T}^2_5 = - \frac{1}{2} t^{3 q} \, W \, {\bf K}_5 + ( y \psi_x - x \psi_y ) {\bf T}_0 \, , \quad {\bf T}^2_6 = -\frac{1}{2} t^{3 q} \, W \, {\bf K}_6 + ( z \psi_x - x \psi_z ) {\bf T}_0 \, , \nonumber \\ & & {\bf T}^2_7 = - \frac{1}{2} t^{3 q} \,  W \, {\bf K}_7 + ( z \psi_y - y \psi_z ) {\bf T}_0 \, , \, \quad {\bf T}^2_8 = - \frac{1}{2} t^{ 3 q} \, W \, {\bf K}^1_8 + [  2(q-1) x \psi - Q_1 ] \, {\bf T}_0  + {\bf f}^2_8 \, , \quad  \nonumber \\ & & {\bf T}^2_9 = - \frac{1}{2} t^{3 q} \, W \, {\bf K}^1_9 + [ 2 (q-1) y \psi - Q_2 ] \, {\bf T}_0 + {\bf f}^2_9 \, , \nonumber \\ & &  {\bf T}^2_{10} = - \frac{1}{2} t^{3 q} \, W \, {\bf K}^1_{10} + [ 2 (q-1) z \psi - Q_3 )] \, {\bf T}_0  + {\bf f}^2_{10} \, ,  \label{c-vectors-i1-1} \\ & & {\bf T}^2_{11,12} = M \, e^{\pm b_1 x + b_2 y } \sin (\ell z ) \, {\bf T}_0 + {\bf f}^2_{11,12} \, , {\bf T}^2_{13,14} = M \, e^{ \pm b_1 x - b_2 y } \sin (\ell z ) \, {\bf T}_0 + {\bf f}^2_{13,14}  \nonumber \\ & & {\bf T}^2_{15,16} = M  \, e^{\pm b_1 x + b_2 y } \cos (\ell z ) \, {\bf T}_0 + {\bf f}^2_{15,16}  \, , {\bf T}^2_{17,18} = M \, e^{\pm b_1 x - b_2 y } \cos (\ell z ) \, {\bf T}_0 + {\bf f}^2_{17,18} \nonumber  \\ & & {\bf T}^2_{19,20} = N \, e^{\pm b_1 x + b_2 y } \sin (\ell z ) \, {\bf T}_0 + {\bf f}^2_{19,20} \, ,  {\bf T}^2_{21,22} = N \, e^{\pm b_1 x - b_2 y } \sin (\ell z ) \, {\bf T}_0 + {\bf f}^2_{21,22}  \nonumber \\ & & {\bf T}^2_{23,24} = N \, e^{\pm b_1 x + b_2 y  } \cos (\ell z ) \, {\bf T}_0 + {\bf f}^2_{23,24}  \, , {\bf T}^2_{25,26} = N \, e^{ \pm b_1 x - b_2 y } \cos (\ell z ) \, {\bf T}_0 + {\bf f}^2_{25,26}  \nonumber
\end{eqnarray}
where $W, {\bf T}_0$ are the same as given in \eqref{W-t0-i}, in which $F=0$ and $A=B=C=t^q$, and $Q_1, Q_2$ and $Q_3$ are defined as the following:
\begin{eqnarray}
  & & Q_1 = 2 (q-1) x \left[ - t \psi_t + (q-1) ( y \psi_y + z \psi_z ) \right] + \left[ (q-1)^2 ( x^2 - y^2 - z^2 ) + t^{2 (1-q)} \right] \psi_x \, , \label{Q1} \\ & & Q_2 = 2 (q-1) y \left[ - t \psi_t + (q-1) ( x \psi_x + z \psi_z ) \right] + \left[ (q-1)^2 ( -x^2 + y^2 - z^2 ) + t^{2 (1-q)} \right] \psi_y \, , \label{Q2} \\ & & Q_3 = 2 (q-1) z \left[ - t \psi_t + (q-1) ( x \psi_x + y \psi_y ) \right] + \left[ (q-1)^2 ( -x^2 - y^2 + z^2 ) + t^{2 (1-q)} \right] \psi_z \, . \label{Q3}
\end{eqnarray}
For some values of $q$ in this subcase, one can obtain different Noether symmetry generators from those above. Now, we will give some examples of this situation, as seen below.

If $L = p = q = 1$ and $F = 0$, then there are again {\it twenty six} Noether symmetries, such that the KVs ${\bf K}_1, {\bf K}_2, {\bf K}_3$, and
\begin{eqnarray}
  & & {\bf X}^3_4 = - t \partial_t + \psi \partial_{\psi} \, , \quad {\bf X}^3_5 = {\bf K}_5 \, , \quad {\bf X}^3_6 = {\bf K}_6 \, , \quad {\bf X}^2_7 = {\bf K}_7 \, , \label{ngs-i1-2-1} \\ & & {\bf X}^3_8 = {\bf K}^2_8 - x \, \psi \, \partial_{\psi} \, , \quad {\bf X}^3_9 = {\bf K}^2_9 - y \, \psi \, \partial_{\psi} \, , \quad {\bf X}^3_{10} = {\bf K}^2_{10} - z \, \psi \, \partial_{\psi} \, , \label{ngs-i1-2-2}  \\ & & {\bf X}^3_{11} =  t^{-1} h_1 (t)e^{ b_1 x + b_2 y} \sin(\ell z) \partial_{\psi} \, , \quad {\bf X}^3_{12} =  t^{-1} h_1 (t) e^{-b_1 x + b_2 y} \sin(\ell z) \partial_{\psi} \, , \label{ngs-i1-2-3}  \\ & & {\bf X}^3_{13} = t^{-1} h_1 (t) e^{ b_1 x - b_2 y} \sin(\ell z) \partial_{\psi} \, , \quad {\bf X}^3_{14} =  t^{-1} h_1 (t) e^{-(b_1 x + b_2 y) } \sin(\ell z) \partial_{\psi} \, ,  \label{ngs-i1-2-4} \\ & & {\bf X}^3_{15} = t^{-1}  h_1 (t) e^{ b_1 x + b_2 y} \cos(\ell z) \partial_{\psi} \, , \quad {\bf X}^3_{16} = t^{-1} h_1 (t) e^{-b_1 x + b_2 y  } \cos(\ell z) \partial_{\psi} \, , \label{ngs-i1-2-5} \\ & & {\bf X}^3_{17} = t^{-1} h_1 (t) e^{ b_1 x - b_2 y} \cos(\ell z) \partial_{\psi} \, , \quad {\bf X}^3_{18} = t^{-1} h_1 (t) e^{-(b_1 x + b_2 y) } \cos(\ell z) \partial_{\psi} \, , \label{ngs-i1-2-6}  \\ & & {\bf X}^3_{19} = t^{-1} h_2 (t) e^{ b_1 x + b_2 y} \sin(\ell z) \partial_{\psi} \, , \quad {\bf X}^3_{20} = t^{-1} h_2 (t) e^{-b_1 x + b_2 y } \sin(\ell z) \partial_{\psi} \, , \label{ngs-i1-2-7} \\ & & {\bf X}^3_{21} = t^{-1}  h_2 (t) e^{ b_1 x - b_2 y} \sin(\ell z) \partial_{\psi} \, , \quad {\bf X}^3_{22} = t^{-1} h_2 (t) e^{-(b_1 x + b_2 y) } \sin(\ell z) \partial_{\psi} \, , \label{ngs-i1-2-8} \\ & & {\bf X}^3_{23} = t^{-1} h_2 (t) e^{ b_1 x + b_2 y} \cos(\ell z) \partial_{\psi} \, , \quad {\bf X}^3_{24} = t^{-1} h_2 (t) e^{-b_1 x + b_2 y } \cos(\ell z) \partial_{\psi} \, , \label{ngs-i1-2-9} \\ & & {\bf X}^3_{25} = t^{-1} h_2 (t) e^{ b_1 x - b_2 y} \cos(\ell z) \partial_{\psi} \, , \quad {\bf X}^3_{26} = t^{-1} h_2 (t) e^{-(b_1 x + b_2 y) } \cos(\ell z) \partial_{\psi} \, , \label{ngs-i1-2-10}
\end{eqnarray}
where ${\bf K}^2_8, {\bf K}^2_9$ and ${\bf K}^2_{10}$ are given by
\begin{eqnarray}
  & & {\bf K}^2_8 = x \, t \, \partial_t + \ln t \, \partial_x \, , \qquad {\bf K}^2_9 = y \, t \, \partial_t + \ln t \, \partial_y \, , \qquad {\bf K}^2_{10} = z \, t \, \partial_t + \ln t \, \partial_z \, , \label{k2-i1-2}
\end{eqnarray}
which are SCKVs with the conformal factors $\sigma = x$ for ${\bf K}^2_8$, $\sigma= y$ for ${\bf K}^2_9$, and $\sigma = z$ for ${\bf K}^2_{10}$. Here, the functions $h_1 (t)$ and $h_2 (t)$ are of the forms $h_1 (t) = t^{\sqrt{1- b_3^2}}$ and $h_2 (t) = t^{-\sqrt{1- b_3^2}}$. The corresponding  gauge vector fields of the above Noether symmetries are
\begin{eqnarray}
  & & {\bf f}^3_8 = -\frac{t}{2} \psi^2 \partial_x \, , \quad {\bf f}^3_9 = -\frac{t}{2} \psi^2 \partial_y \, , \quad {\bf f}^3_{10} = -\frac{t}{2} \psi^2 \partial_z \, , \nonumber
\\
  & & {\bf f}^3_{11,12} = h_1 (t) e^{ \pm b_1 x + b_2 y} \sin (\ell z) \psi \left( (1- \sqrt{1 -b_3^2} ) t , \pm b_1 , b_2 , \ell \cot (\ell z) \right) \, , \nonumber
\\
  & & {\bf f}^3_{13,14} = h_1 (t) e^{ \pm b_1 x - b_2 y} \sin (\ell z) \psi \left( (1- \sqrt{1 -b_3^2} ) t ,\pm  b_1 , - b_2 , \ell \cot (\ell z) \right) \,  \nonumber
\\
  & & {\bf f}^3_{15,16} = h_1 (t) e^{ \pm  b_1 x + b_2 y } \cos (\ell z) \psi \left( (1- \sqrt{1 -b_3^2} ) t , \pm b_1 , b_2 , - \ell \tan (\ell z) \right) \,  , \nonumber
\\
  & & {\bf f}^3_{17,18} = h_1 (t) e^{ \pm b_1 x - b_2 y } \cos (\ell z) \psi \left( (1- \sqrt{1 -b_3^2} ) t , \pm b_1 , -b_2 , - \ell \tan (\ell z) \right) \, , \label{gauge-v-i1-2}
\\
  & & {\bf f}^3_{19,20} = h_2 (t) e^{ \pm b_1 x + b_2 y } \sin (\ell z) \psi \left( (1- \sqrt{1 -b_3^2} ) t , \pm b_1 , b_2 ,  \ell \cot (\ell z) \right) \, , \nonumber
\\
  & & {\bf f}^3_{21,22} = h_2 (t) e^{ \pm b_1 x - b_2 y } \sin (\ell z) \psi \left( (1- \sqrt{1 -b_3^2} ) t , \pm b_1 ,-b_2 ,  \ell \cot (\ell z) \right) \, , \nonumber
\\
  & & {\bf f}^3_{23,24} = h_2 (t) e^{ \pm b_1 x + b_2 y } \cos (\ell z) \psi \left( (1- \sqrt{1 -b_3^2} ) t , \pm b_1 , b_2 ,  -\ell \tan (\ell z) \right) \, , \nonumber
\\
  & & {\bf f}^3_{25,26} = h_2 (t) e^{ \pm b_1 x - b_2 y } \cos (\ell z) \psi \left( (1- \sqrt{1 -b_3^2} ) t , \pm b_1 , - b_2 ,  -\ell \tan (\ell z) \right) \, . \nonumber
\end{eqnarray}

Using the conservation law relation \eqref{c-vector}, the conserved flow vectors for ${\bf X}^3_4,...,{\bf X}^3_{26}$ become
\begin{eqnarray}
  & &  {\bf T}^3_4 = \frac{1}{2} t^3 \, W \, \partial_t + ( t \psi_t  + \psi ) {\bf T}_0  \, , \qquad {\bf T}^3_5 = - \frac{1}{2} t^3 \, W \, {\bf K}_5 + ( x \psi_y - y \psi_x ) {\bf T}_0 \, , \nonumber \\ & & {\bf T}^3_6 = - \frac{1}{2} t^3 \, W \, {\bf K}_6 + ( x \psi_z - \psi_x ) {\bf T}_0 \, , \quad {\bf T}^3_7 = - \frac{1}{2} t^3 \, W \, {\bf K}_7 + ( y \psi_z - z \psi_y ) {\bf T}_0 \, , \nonumber \\ & & {\bf T}^3_8 = - \frac{1}{2} t^3 \, W \, {\bf K}^2_8 - \left( x ( t \psi_t + \psi ) + \ln t \, \psi_x \right) {\bf T}_0  + {\bf f}^3_8  \, , \nonumber \\ & & {\bf T}^3_9 = - \frac{1}{2} t^3 \, W \, {\bf K}^2_9 - \left( y ( t \psi_t + \psi ) + \ln t \, \psi_y \right) {\bf T}_0  + {\bf f}^3_9  \, , \label{c-vectors-i1-2} \\ & &  {\bf T}^3_{10} = - \frac{1}{2} t^3 \, W \, {\bf K}^2_{10} - \left( z ( t \psi_t + \psi ) + \ln t \, \psi_z \right) {\bf T}_0 + {\bf f}^3_{10}  \, , \nonumber  \\ & & {\bf T}^3_{11,12} =  t^{-1} h_1 (t) e^{ \pm b_1 x + b_2 y} \sin (\ell z) {\bf T}_0 + {\bf f}^3_{11,12} \, , \,\, {\bf T}^3_{13,14} = t^{-1} h_1 (t) e^{ \pm b_1 x - b_2 y} \sin (\ell z) {\bf T}_0 + {\bf f}^3_{13,14}  \nonumber \\ & & {\bf T}^3_{15,16} = t^{-1} h_1 (t) e^{ \pm b_1 x + b_2 y} \cos (\ell z) {\bf T}_0 + {\bf f}^3_{15,16} \, , \,\, {\bf T}^3_{17,18} = t^{-1} h_1 (t) e^{ \pm b_1 x - b_2 y} \cos (\ell z) {\bf T}_0 + {\bf f}^3_{17,18}  \nonumber \\ & & {\bf T}^3_{19,20} = t^{-1} h_2 (t) e^{ \pm b_1 x + b_2 y} \sin (\ell z) {\bf T}_0 + {\bf f}^3_{19,20} \, , \,\, {\bf T}^3_{21,22} = t^{-1} h_2 (t) e^{ \pm b_1 x - b_2 y} \sin (\ell z) {\bf T}_0 + {\bf f}^3_{21,22}  \nonumber \\ & & {\bf T}^3_{23,24} = t^{-1} h_2 (t) e^{ \pm b_1 x + b_2 y} \cos (\ell z) {\bf T}_0 + {\bf f}^3_{23,24} \, , \,\, {\bf T}^3_{25,26} = t^{-1} h_2 (t) e^{ \pm b_1 x - b_2 y} \cos (\ell z) {\bf T}_0 + {\bf f}^3_{25,26}  \nonumber
\end{eqnarray}
where ${\bf T}_0$ and $W$ are the same as given in \eqref{W-t0-i} by taking $A = B = C = t$ and $F = 0$.

When $L = p= q = 1/2$ and $F = 0$, we obtain {\it thirty one} Noether symmetries, which is the maximum number of symmetries, and these include six KVs ${\bf K}_1, {\bf K}_2, {\bf K}_3, {\bf K}_5, {\bf K}_6, {\bf K}_7$ and the following:
\begin{eqnarray}
  & &  {\bf X}^4_{7} = {\bf X}^2_4 \, , \quad {\bf X}^4_{8} = {\bf X}^2_8 \, , \quad {\bf X}^4_{9} = {\bf X}^2_9 \, ,  \quad  {\bf X}^4_{10} = {\bf X}^2_{10} \, , \label{ngs-i1-3-1}  \\ & & {\bf X}^4_{11} = {\bf K}_{11} - \frac{\psi}{2 \sqrt{t} } \partial_{\psi} \, , \quad {\bf X}^4_{12} = {\bf K}_{12} - \frac{x \, \psi}{2 \sqrt{t}} \partial_{\psi} \, ,  \quad  {\bf X}^4_{13} = {\bf K}_{13} - \frac{y \, \psi}{2 \sqrt{t}} \partial_{\psi} \, , \label{ngs-i1-3-2}  \\  & & {\bf X}^4_{14} = {\bf K}_{14} - \frac{z \, \psi}{2 \sqrt{t}} \partial_{\psi} \, , \quad  {\bf X}^4_{15} =  {\bf K}_{15} - \frac{1}{2 \sqrt{t}} \Big{[} 3 t + \frac{1}{4} (x^2 + y^2 + z^2) \Big{]} \psi \partial_{\psi} \, , \qquad \label{ngs-i1-3-3} \\ & & {\bf X}^4_{16} = h_3 (t) e^{ b_1 x + b_2 y} \sin (\beta z) \partial_{\psi} \, , \quad  {\bf X}^4_{17} = h_3 (t) e^{ - b_1 x + b_2 y} \sin (\beta z) \partial_{\psi} \, , \label{ngs-i1-3-4} \\ & & {\bf X}^4_{18} = h_3 (t) e^{ b_1 x - b_2 y} \sin (\beta z) \partial_{\psi} \, , \quad  {\bf X}^4_{19} = h_3 (t) e^{ -( b_1 x + b_2 y ) } \sin (\beta z) \partial_{\psi} \, , \label{ngs-i1-3-5} \\ & & {\bf X}^4_{20} = h_7 (t) e^{ b_1 x + b_2 y} \cos (\beta z) \partial_{\psi} \, , \quad  {\bf X}^4_{21} = h_3 (t) e^{ - b_1 x + b_2 y} \cos (\beta z) \partial_{\psi} \, , \label{ngs-i1-3-6} \\ & & {\bf X}^4_{22} = h_3 (t) e^{ b_1 x - b_2 y} \cos (\beta z) \partial_{\psi} \, , \quad  {\bf X}^4_{23} = h_3 (t) e^{ -( b_1 x + b_2 y ) } \cos (\beta z) \partial_{\psi} \, , \label{ngs-i1-3-7} \\ & & {\bf X}^4_{24} = h_4 (t) e^{ b_1 x + b_2 y} \sin (\beta z) \partial_{\psi} \, , \quad  {\bf X}^4_{25} = h_4 (t) e^{ - b_1 x + b_2 y} \sin (\beta z) \partial_{\psi} \, , \label{ngs-i1-3-8} \\ & & {\bf X}^4_{26} = h_4 (t) e^{ b_1 x - b_2 y} \sin (\beta z) \partial_{\psi} \, , \quad  {\bf X}^4_{27} = h_4 (t) e^{ -( b_1 x + b_2 y ) } \sin (\beta z) \partial_{\psi} \, , \label{ngs-i1-3-9} \\ & & {\bf X}^4_{28} = h_4 (t) e^{ b_1 x + b_2 y} \cos (\beta z) \partial_{\psi} \, , \quad  {\bf X}^4_{29} = h_4 (t) e^{ - b_1 x + b_2 y} \cos (\beta z) \partial_{\psi} \, , \label{ngs-i1-3-10} \\ & & {\bf X}^4_{30} = h_4 (t) e^{ b_1 x - b_2 y} \cos (\beta z) \partial_{\psi} \, , \quad  {\bf X}^4_{31} = h_4 (t) e^{ -( b_1 x + b_2 y ) } \cos (\beta z) \partial_{\psi} \, , \label{ngs-i1-3-11}
\end{eqnarray}
where $\beta = \sqrt{b_1^2 + b_2^2 - b_3^2}$, $ h_3 (t) = \sinh (2 b_3 \sqrt{t} ) / \sqrt{t}$, and $ h_4 (t) = \cosh (2 b_3 \sqrt{t} ) / \sqrt{t}$. Furthermore, ${\bf K}_{11}$, ${\bf K}_{12}$, ${\bf K}_{13}$, ${\bf K}_{14}$, and ${\bf K}_{15}$ are defined as follows:
\begin{eqnarray}
  & & {\bf K}_{11} = \sqrt{t} \, \partial_t \, , \,\, {\bf K}_{12} = \sqrt{t} ( x \partial_t + 2 \partial_x )  \, , \,\, {\bf K}_{13} = \sqrt{t} ( y \partial_t + 2 \partial_y ) \, , \,\,  {\bf K}_{14} = \sqrt{t} ( z \partial_t + 2 \partial_z ) \, , \qquad \label{ckv-i1-3-1}\\ & & {\bf K}_{15} = \sqrt{t} \left[ \Big{(} t + \frac{1}{4} (x^2 + y^2 + z^2) \Big{)} \partial_t + x \partial_x + y \partial_y + z \partial_z \right]  \, , \label{ckv-i1-3-2}
\end{eqnarray}
which are CKVs of Bianchi I spacetime and the conformal factors are $\sigma = 1/(2 \sqrt{t})$ for ${\bf K}_{11}$, $\sigma = x /(2 \sqrt{t})$ for ${\bf K}_{12}$, $\sigma = y /(2 \sqrt{t})$ for ${\bf K}_{13}$, $\sigma = z /(2 \sqrt{t})$ for ${\bf K}_{14}$, and $\sigma = [ 3 t + (x^2 + y^2 + z^2)/4] / \sqrt{t}$ for ${\bf K}_{15}$. Here, the gauge vector fields for the Noether symmetries obtained in \eqref{ngs-i1-3-1}--\eqref{ngs-i1-3-11} are
\begin{eqnarray}
  & & {\bf f}^4_{8} = \frac{1}{2} \sqrt{t}\, \psi^2 \partial_x \, , \quad {\bf f}^4_{9} = \frac{1}{2} \sqrt{t} \, \psi^2 \partial_y \, , \quad {\bf f}^4_{10} = \frac{1}{2} \sqrt{t} \, \psi^2 \partial_z \, ,  \quad {\bf f}^4_{11} = -\frac{1}{8} \psi^2 \partial_t \, , \nonumber \\ & & {\bf f}^4_{12} = - \frac{1}{8} \psi^2 ( x \partial_t + 2 \partial_x  ) \, , \quad {\bf f}^4_{13} = - \frac{1}{8} \psi^2 ( y \partial_t + 2 \partial_y  ) \, ,  \quad {\bf f}^4_{14} = - \frac{1}{8} \psi^2 ( z \partial_t + 2 \partial_z  ) \, , \nonumber \\ & & {\bf f}^4_{15} = - \frac{1}{8} \psi^2 \left( \Big{[} 3\, t - \frac{1}{4} (x^2 + y^2 + z^2 ) \Big{]} \partial_t + x \partial_x + y \partial_y + z \partial_z \right) \,  ,  \nonumber \\ & & {\bf f}^4_{16,17} = \sqrt{t}\, h_3 (t) e^{ \pm b_1 + b_2 y} \sin (\beta z) \psi \left( \frac{1}{2} -b_3 \sqrt{t} \coth(2 b_3 \sqrt{t}) , \pm b_1 , b_2 , \beta \cot ( \beta z) \right)  \, , \nonumber \\ & & {\bf f}^4_{18,19} = \sqrt{t} \, h_3 (t) e^{ \pm b_1 -  b_2 y} \sin (\beta z) \psi \left( \frac{1}{2} - b_3 \sqrt{t} \coth(2 b_3 \sqrt{t}) , \pm b_1 , -b_2 , \beta \cot ( \beta z) \right)  \, , \nonumber \\ & & {\bf f}^4_{20,21} = \sqrt{t}\, h_3 (t) e^{ \pm b_1 + b_2 y} \cos (\beta z) \psi \left( \frac{1}{2} - b_3 \sqrt{t} \coth(2 b_3 \sqrt{t}) , \pm b_1 , b_2 , -\beta \tan (\beta z) \right)  \, , \\ & & {\bf f}^4_{22,23} = \sqrt{t}\, h_3 (t) e^{ \pm b_1 - b_2 y} \cos (\beta z) \psi \left( \frac{1}{2} - b_3 \sqrt{t} \coth(2 b_3 \sqrt{t}) , \pm b_1 , -b_2 , -\beta \tan (\beta z) \right)  \, , \nonumber \\ & & {\bf f}^4_{24,25} = \sqrt{t}\, h_4 (t) e^{ \pm b_1 + b_2 y} \sin (\beta z) \psi \left( \frac{1}{2} -b_3 \sqrt{t} \coth(2 b_3 \sqrt{t}) , \pm b_1 , b_2 , \beta \cot ( \beta z) \right)  \, , \nonumber \\ & & {\bf f}^4_{26,27} = \sqrt{t}\, h_4 (t) e^{ \pm b_1 - b_2 y} \sin (\beta z) \psi \left( \frac{1}{2} -b_3 \sqrt{t} \coth(2 b_3 \sqrt{t}) , \pm b_1 , -b_2 , \beta \cot ( \beta z) \right)  \, , \nonumber \\ & & {\bf f}^4_{28,29} = \sqrt{t}\, h_4 (t) e^{ \pm b_1 + b_2 y} \cos (\beta z) \psi \left( \frac{1}{2} -b_3 \sqrt{t} \coth(2 b_3 \sqrt{t}) , \pm b_1 , b_2 , -\beta \tan ( \beta z) \right)  \, , \nonumber \\ & & {\bf f}^4_{30,31} = \sqrt{t}\, h_4 (t) e^{ \pm b_1 - b_2 y} \cos (\beta z) \psi \left( \frac{1}{2} - b_3 \sqrt{t} \coth(2 b_3 \sqrt{t}) , \pm b_1 , -b_2 , -\beta \tan ( \beta z) \right)  \, . \nonumber
\end{eqnarray}
Then, the corresponding conserved flow vectors for the Noether symmetries given in \eqref{ngs-i1-3-1}--\eqref{ngs-i1-3-11} yield
\begin{eqnarray}
  & & {\bf T}^4_{8} = {\bf T}^2_8 \,\,\, {\rm with} \,\, q=1/2 \, , \qquad {\bf T}^4_{9} = {\bf T}^2_9 \,\,\, {\rm with} \,\, q=1/2 \, , \nonumber \\ & & {\bf T}^4_{10} = {\bf T}^2_{10} \,\,\, {\rm with} \,\, q=1/2 \, ,  \quad {\bf T}^4_{11} = - \frac{1}{2} t^{3/2} \, W \, {\bf K}_{11} - \sqrt{t} \Big{(} \psi_t + \frac{\psi}{2\, t}  \Big{)} {\bf T}_0 + {\bf f}^4_{11} \, , \nonumber \\ & & {\bf T}^4_{12} = - \frac{1}{2} t^{3/2} \, W \, {\bf K}_{12} - \sqrt{t} \Big{(} x ( \psi_t + \frac{\psi}{2\, t} ) + 2 \, \psi_x \Big{)} {\bf T}_0  + {\bf f}^4_{12}  \, , \nonumber \\ & &  {\bf T}^4_{13} = - \frac{1}{2} t^{3/2} \, W \, {\bf K}_{13} - \sqrt{t} \Big{(} y ( \psi_t + \frac{\psi}{2\, t} ) + 2 \psi_y \Big{)} {\bf T}_0  + {\bf f}^4_{13} \, ,  \nonumber \\ & &  {\bf T}^4_{14} = - \frac{1}{2} t^{3/2} \, W \, {\bf K}_{14} - \sqrt{t} \Big{(} z ( \psi_t + \frac{\psi}{2\, t} ) + 2 \psi_z \Big{)} {\bf T}_0 + {\bf f}^4_{14}  \, , \label{c-vectors-ii-3} \\ & & {\bf T}^4_{15} = - \frac{1}{2} t^{3/2} \, W \, {\bf K}_{15} - \sqrt{t} \Big{[} Q_4 + \frac{1}{2 \, t} \Big{(} 3 \, t +  \frac{1}{4}(x^2 + y^2 + z^2) \Big{)} \psi \Big{]} {\bf T}_0 + {\bf f}^4_{15} \, ,  \nonumber  \\ & & {\bf T}^4_{16,17} =  h_3 (t) e^{\pm b_1 x  + b_2 y} \sin (\beta z) {\bf T}_0 + {\bf f}^4_{16,17} \, , \,\, {\bf T}^4_{18,19} = h_3 (t) e^{\pm b_1 x  - b_2 y} \sin (\beta z) {\bf T}_0 + {\bf f}^4_{18,19}  \nonumber \\ & &  {\bf T}^4_{20,21} =  h_3 (t) e^{\pm b_1 x  + b_2 y} \cos (\beta z) {\bf T}_0 + {\bf f}^4_{20,21} \, , \,\, {\bf T}^4_{22,23} =  h_3 (t) e^{\pm b_1 x  - b_2 y} \cos (\beta z) {\bf T}_0 + {\bf f}^4_{22,23} \nonumber \\ & & {\bf T}^4_{24,25} =  h_4 (t) e^{\pm b_1 x  + b_2 y} \sin (\beta z) {\bf T}_0 + {\bf f}^4_{24,25} \, , \,\, {\bf T}^4_{26,27} = h_4 (t) e^{\pm b_1 x  - b_2 y} \sin (\beta z) {\bf T}_0 + {\bf f}^4_{26,27}  \nonumber \\ & & {\bf T}^4_{28,29} = h_4 (t) e^{\pm b_1 x  + b_2 y} \cos (\beta z) {\bf T}_0 + {\bf f}^4_{28,29} \, , \,\, {\bf T}^4_{30,31} = h_4 (t) e^{\pm b_1 x  - b_2 y} \cos (\beta z) {\bf T}_0 + {\bf f}^4_{30,31}  \nonumber
\end{eqnarray}
where ${\bf T}_0$ and $W$ are given in \eqref{W-t0-i} through $F=0, A = B = C = t^{1/2}$, and $Q_4$ is defined by
\begin{eqnarray}
  & & Q_4 = \left[ t + \frac{1}{4} (x^2 + y^2 + z^2) \right] \psi_t + x \psi_x + y \psi_y + z \psi_z  \, . \label{Q4}
\end{eqnarray}

\noindent {\bf Subcase (ii.3)}. If $A(t) = B(t) = C(t) = t^q$ through $F= (b \psi + c)^4$ (see Ref. \cite{camci2014} for selection of this function to study the Noether symmetries in the flat FLRW spacetime), where $b$ and $c$ are constants, then we find that there are {\it ten} Noether symmetries: \begin{eqnarray}
    & & {\bf X}^6_1 = {\bf K}_1 \, ,  \quad {\bf X}^6_2 = {\bf K}_2 \, , \quad {\bf X}^6_3 = {\bf K}_3 \, , \quad {\bf X}^6_4 = {\bf K}_4 + \frac{1}{b} ( b \psi + c) \partial_{\psi} \, ,  \\ & & {\bf X}^6_5 =  {\bf K}_5 \, , \quad {\bf X}^6_6 = {\bf K}_6 \, , \quad {\bf X}^6_7 = {\bf K}_7 \, , \\ & & {\bf X}^6_8 = {\bf K}^1_8 + \frac{2}{b} (q-1) ( b \psi + c) \, x\, \partial_{\psi},  \quad \,\,\, {\rm with\,\, } \quad  {\bf f}^6_8 = \frac{1}{b} (q-1) t^q \psi ( b \psi + c) \partial_x \, ,  \\ & & {\bf X}^6_9 = {\bf K}^1_9 + \frac{2}{b} (q-1) ( b \psi + c) \, y \, \partial_{\psi},  \quad \,\,\, {\rm with\,\, } \quad {\bf f}^6_9 = \frac{1}{b} (q-1) t^q \psi ( b \psi + c) \partial_y \, ,  \\ & & {\bf X}^6_{10} = {\bf K}^1_{10} + \frac{2}{b} (q-1) ( b \psi + c) \, z \, \partial_{\psi},  \quad {\rm with\,\, }  \quad {\bf f}^6_{10} = \frac{1}{b} (q-1) t^q \psi ( b \psi + c) \partial_z \, ,
\end{eqnarray}
where ${\bf K}^1_8, {\bf K}^1_9$, and ${\bf K}^1_{10}$ are the same as given in \eqref{sckv-i1-1-8}, \eqref{sckv-i1-1-9}, and \eqref{sckv-i1-1-10}, respectively. Then, the conserved vectors of ${\bf X}^6_4,...,{\bf X}^6_{10}$ are
\begin{eqnarray}
  & & {\bf T}^6_4 = - \frac{1}{2} t^{3 q} \, W \, {\bf K}_4 + \Big{(} \psi + \frac{c}{b} - Q \Big{)} {\bf T}_0  \, , \quad {\bf T}^6_5 = {\bf T}^2_5 \, , \quad {\bf T}^6_6 = {\bf T}^2_6 \, , \quad {\bf T}^6_7 = {\bf T}^2_7 \, , \nonumber \\ & & {\bf T}^6_8 = - \frac{1}{2} t^{3 q} \, W \, {\bf K}^1_8 + \left[ \frac{2}{b} (q-1) x ( b \psi + c) - Q_1 \right] {\bf T}_0  + {\bf f}^6_8 \, , \nonumber \\ & & {\bf T}^6_9 = - \frac{1}{2} t^{3 q} \, W \, {\bf K}^1_9 + \left[ \frac{2}{b} (q-1) y ( b \psi + c) - Q_2 \right] {\bf T}_0  + {\bf f}^6_9 \, , \\ & & {\bf T}^6_{10} = - \frac{1}{2} t^{3 q} \, W \, {\bf K}^1_{10} + \left[ \frac{2}{b} (q-1) z ( b \psi + c) - Q_3 \right] {\bf T}_0 + {\bf f}^6_{10} \, , \nonumber
\end{eqnarray}
where $W = -\psi_t^2 + t^{-2 q} ( \psi_x^2 + \psi_y^2 +  \psi_z^2 ) - 2 ( b \psi + c)^4$,  $Q = - t \psi_t + (q-1) ( x \psi_x + y \psi_y + z \psi_z )$, $Q_i$'s $(i=1,2,3)$ are the same as in \eqref{Q1}--\eqref{Q3}, and $q \neq 0, 1, 1/2$.
For certain values of $q$, one can obtain either {\it ten} or {\it fifteen} Noether symmetry generators. Now, we will provide some examples illustrating this situation below.

\noindent {\bf (ii.3.1)}: For $q = 1$, there are again {\it ten} Noether symmetries, such that ${\bf X}^7_1 = {\bf K}_1$, $ {\bf X}^7_2 = {\bf K}_2$, $ {\bf X}^7_3= {\bf K}_3$ and 
\begin{eqnarray}
    & & {\bf X}^7_4 = - t \, \partial_t + \frac{1}{b} ( b \psi + c) \, \partial_{\psi} \, , \quad  {\bf X}^7_5 = {\bf K}_5 \, , \quad {\bf X}^7_6 = {\bf K}_6 \, , \quad {\bf X}^7_7 = {\bf K}_7 \, , \label{ngs-i3-1-1}  \\ & & {\bf X}^7_8 = {\bf K}^2_8 - \frac{x}{b} ( b \psi + c) \, \partial_{\psi} \quad {\rm with} \quad {\bf f}^7_8 = - \frac{t}{2 b} \psi ( b \psi + c) \, \partial_x  \, , \label{ngs-i3-1-2} \\ & & {\bf X}^7_9 = {\bf K}^2_9 - \frac{y}{b} ( b \psi + c) \, \partial_{\psi} \quad {\rm with} \quad {\bf f}^7_9 = - \frac{t}{2 b} \psi ( b \psi + c) \, \partial_y \, , \label{ngs-i3-1-3} \\ & & {\bf X}^7_{10} = {\bf K}^2_{10} - \frac{z}{b} ( b \psi + c) \, \partial_{\psi} \quad {\rm with} \quad {\bf f}^7_{10} = - \frac{t}{2 b} \psi ( b \psi + c) \,  \partial_z  \, , \label{ngs-i3-1-4}
\end{eqnarray}
where ${\bf K}^2_8, {\bf K}^2_9, {\bf K}^2_{10}$ are the same vector fields defined in \eqref{k2-i1-2}. The conserved flow vectors for the above Noether symmetries are
\begin{eqnarray}
  & & {\bf T}^7_5 = {\bf T}^3_5 \, , \quad {\bf T}^7_6 = {\bf T}^3_6 \, , \quad {\bf T}^7_7 = {\bf T}^3_7 \, , \quad  {\bf T}^7_8 = - \frac{1}{2} t^3 \, W \, {\bf K}^2_8 - \Big{[} x ( t \psi_t + \frac{1}{b} ( b \psi + c) ) + \ln t \, \psi_x \Big{]} {\bf T}_0  + {\bf f}^7_8  \, , \nonumber \\ & & {\bf T}^7_9 = - \frac{1}{2} t^3 \, W \, {\bf K}^2_9 - \Big{[} y ( t \psi_t + \frac{1}{b} ( b \psi + c) ) + \ln t \, \psi_y \Big{]} {\bf T}_0 + {\bf f}^7_9  \, ,   \\ & &  {\bf T}^7_{10} = - \frac{1}{2} t^3 \, W \, {\bf K}^2_{10} - \Big{[} z ( t \psi_t + \frac{1}{b} ( b \psi + c) ) + \ln t \, \psi_z \Big{]} {\bf T}_0  + {\bf f}^7_{10}  \, ,\nonumber
\end{eqnarray}
where  ${\bf T}^3_5, {\bf T}^3_6, {\bf T}^3_7$ are of the same form as obtained in \eqref{c-vectors-i1-2}, and $W = -\psi_t^2 + t^{-2} ( \psi_x^2 + \psi_y^2 + \psi_z^2 ) - (b \psi + c)^4$.

\noindent {\bf (ii.3.2)}: For $q = 1/2$, we find that the number of Noether symmetries is {\it fifteen}, which are the KVs ${\bf K}_1, {\bf K}_2, {\bf K}_3, {\bf K}_5, {\bf K}_6, {\bf K}_7$ and
\begin{eqnarray}
    & & {\bf X}^8_4  =  {\bf K}_4 +  \frac{1}{b} ( b \psi + c ) \partial_{\psi} \, , \,\, \quad {\bf X}^8_8 = {\bf K}^1_8 - \frac{x}{b} ( b \psi + c) \partial_{\psi}  \, ,   \label{ngs-i3-2-1}  \\ & &  {\bf X}^8_9 = {\bf K}^1_9 - \frac{y}{b} ( b \psi + c) \partial_{\psi} \, , \quad  \,\, {\bf X}^8_{10} = {\bf K}^1_{10} - \frac{z}{b} ( b \psi + c) \partial_{\psi} \, ,  \label{ngs-i3-2-2} \\ & & {\bf X}^8_{11} = {\bf K}_{11} - \frac{( b \psi + c) }{2 \, b \sqrt{t} } \partial_{\psi} \, , \quad {\bf X}^8_{12} = {\bf K}_{12} - \frac{x \, ( b \psi+ c) }{2 \, b \sqrt{t}} \partial_{\psi} \, ,  \,\,  {\bf X}^8_{13} = {\bf K}_{13} - \frac{y \, ( b \psi + c)}{2 \sqrt{t}} \partial_{\psi} \, , \label{ngs-i3-2-3}  \\  & & {\bf X}^8_{14} = {\bf K}_{14} - \frac{z \, ( b \psi + c)}{2\,  b \sqrt{t}} \partial_{\psi} \, , \quad  {\bf X}^8_{15} =  {\bf K}_{15} - \frac{1}{2 \, b \sqrt{t}} \Big{[} 3 t + \frac{1}{4} (x^2 + y^2 + z^2) \Big{]} ( b \psi + c) \partial_{\psi} \, , \qquad \label{ngs-i3-2-4}
\end{eqnarray}
where ${\bf K}_{11},...,{\bf K}_{15}$ are of the form given in Equations \eqref{ckv-i1-3-1} and \eqref{ckv-i1-3-2}. Then, one can obtain the corresponding non-zero gauge vectors, such that
\begin{eqnarray}
  & & {\bf f}^8_8 =- \frac{\sqrt{t}}{2\, b} \, \psi ( b \psi + 2 \, c) \partial_x \, , \quad {\bf f}^8_9 = - \frac{\sqrt{t}}{2 \, b}  \, \psi ( b \psi + 2 \, c) \partial_y \, , \quad {\bf f}^8_{10} = - \frac{\sqrt{t}}{2 \, b} \, \psi ( b \psi + 2 \, c ) \partial_z \, , \nonumber \\ & & {\bf f}^8_{11} = - \frac{1}{8\, b}  \psi ( b \psi + 2 \, c ) \partial_x \, , \qquad {\bf f}^8_{12} = - \frac{1}{8 \, b}  \psi ( b \psi + 2 \, c ) ( x \partial_t + 2 \, \partial_x ) \, , \nonumber \\ & & {\bf f}^8_{13} = - \frac{1}{8 \, b}  \psi ( b \psi + 2 \, c ) ( y \partial_t + 2 \, \partial_y ) \, , \quad {\bf f}^8_{14} = - \frac{1}{8 \, b}  \psi ( b \psi + 2 \, c ) ( z \partial_t + 2 \, \partial_z ) \, ,  \\ & & {\bf f}^8_{15} = - \frac{1}{4 \, b}  \psi ( b \psi + 2 \, c ) \left[ \left( -3 t + \frac{1}{4} (x^2 + y^2 + z^2 ) \right) +  x \partial_x + y \partial_y + z \partial_z \right] \, . \nonumber
\end{eqnarray}

The conserved vectors for the Noether symmetries obtained in \eqref{ngs-i3-2-1}--\eqref{ngs-i3-2-4} yield
\begin{eqnarray}
  & & {\bf T}^8_{4} = - \frac{1}{2} t^{3/2} \, W \, {\bf K}_{4} +  \Big{(} \psi  + \frac{c}{b} - Q \Big{)} {\bf T}_0  \, , \quad {\bf T}^8_{8} = - \frac{1}{2} t^{3/2} \, W \, {\bf K}^1_{8} -  \Big{(} Q_1 + \frac{x}{b} (b \, \psi + c)  \Big{)} {\bf T}_0  + {\bf f}^8_{8} \, , \nonumber \\ & &  {\bf T}^8_{9} = - \frac{1}{2} t^{3/2} \, W \, {\bf K}^1_{9} - \Big{(} Q_2 + \frac{y}{b} (b \, \psi + c)  \Big{)} {\bf T}_0  + {\bf f}^8_{19} \, , \nonumber \\ & & {\bf T}^8_{10} = - \frac{1}{2} t^{3/2} \, W \, {\bf K}^1_{10} - \Big{(} Q_3 + \frac{z}{b} (b \, \psi + c)  \Big{)} {\bf T}_0  + {\bf f}^8_{10} \, ,  \nonumber \\ & &  {\bf T}^4_{11} = - \frac{1}{2} t^{3/2} \, W \, {\bf K}_{11} - \sqrt{t} \Big{(} \psi_t + \frac{\psi}{2\, t}  \Big{)} {\bf T}_0 + {\bf f}^4_{11} \, , \label{c-vectors-i3-2} \\ & & {\bf T}^8_{12} = - \frac{1}{2} t^{3/2} \, W \, {\bf K}_{12} - \sqrt{t} \Big{[} x \Big{(} \psi_t + \frac{1}{2\, b \, t} ( b \, \psi + c) \Big{)} + 2 \, \psi_x \Big{]} {\bf T}_0  + {\bf f}^8_{12}  \, , \nonumber \\ & &  {\bf T}^8_{13} = - \frac{1}{2} t^{3/2} \, W \, {\bf K}_{13} - \sqrt{t} \Big{[} y \Big{(} \psi_t + \frac{1}{2\, b \, t} ( b \, \psi + c) \Big{)} + 2 \, \psi_y \Big{]} {\bf T}_0 + {\bf f}^8_{13} \, ,  \nonumber \\ & &  {\bf T}^8_{14} = - \frac{1}{2} t^{3/2} \, W \, {\bf K}_{14} - \sqrt{t} \Big{[} z \Big{(} \psi_t + \frac{1}{2\, b \, t} ( b \, \psi + c) \Big{)} + 2 \, \psi_z \Big{]} {\bf T}_0  + {\bf f}^8_{14}  \, , \nonumber \\ & & {\bf T}^8_{15} = - \frac{1}{2} t^{3/2} \, W \, {\bf K}_{15} - \sqrt{t} \Big{[} Q_4 + \frac{1}{2 \, b\, t} \Big{(} 3 \, t +  \frac{1}{4}(x^2 + y^2 + z^2) \Big{)} (b \, \psi + c) \Big{]} {\bf T}_0  + {\bf f}^8_{15} \, ,  \nonumber
\end{eqnarray}
where $Q_0, Q_1, Q_2, Q_3$, and $Q_4$ are given in Equations \eqref{Q-i1}, \eqref{Q1}, \eqref{Q2}, \eqref{Q3}, and \eqref{Q4}, respectively.

%4
\section{The Field Equations}
\label{sec:4}

In this section, we briefly examine the gravitational field equations of the Bianchi I spacetime when influenced by an imperfect fluid source. The general expression for the energy-momentum tensor of an imperfect fluid is given by
\begin{equation}
  T_{i j} = ( \rho + P ) u_i u_j + P \, g_{ij}  + ( P_x - P) \chi_i \chi_j + (P_y - P) y_i y_j + (P_z - P) z_i z_j  \, , \label{emt}
\end{equation}
where $\rho$ is the energy density; and $P_x, P_y$, and $P_z$ are the anisotropic pressures in the $x, y$, and $z$ directions, respectively. $P$ represents the isotropic pressure and is defined as $P=(P_x + P_y + P_z)/3$. The energy density $\rho$ is measured using a comoving observer with the timelike unit four-velocity $u^i$, satisfying the normalization condition $u^i u_i = -1$. Additionally, $\chi^i, y^i, z^i$ are spacelike vectors with orthonormality conditions: $\chi^i \chi_i = 1$, $y^i y_i = 1$, $z^i z_i = 1$, $\chi^i y_i = 0$, $\chi^i z_i = 0$, $y^i z_i =0$, $\chi^i u_i = 0$, $y^i u_i = 0$ and $z^i u_i = 0$. When $P_x = P_y = P_z = P$,  the energy-momentum tensor $T_{i j}$ for the imperfect fluid, as given in \eqref{emt}, reduces to that of a perfect fluid.
The pressure is treated as a vectorial quantity, consistent with the nature of anisotropic fluids. The equation of state (EoS) parameter for the imperfect fluid in the Bianchi I spacetime may be determined separately along each spatial axis, denoted as $w_i = P_i / \rho$, where $P_i = \{ P_x, P_y, P_z \}$. The average EoS parameter for the imperfect fluid in the Bianchi I spacetime is defined as $w = P/ \rho = ( w_x + w_y + w_z) /3$. A well-known example of dark energy occurs within the range $w < -\frac{1}{3}$, indicating an accelerating phase for the universe. This form of dark energy is termed quintessence when $-\frac{1}{3} > w > -1$, and labeled a phantom fluid if $w < -1$.

In the geometry described by the Bianchi I metric given in \eqref{b1}, the timelike four-velocity in comoving coordinates is $u^i = \delta^i_0$. From the properties of the spacelike vectors, it follows that $\chi^i = A^{-1} \delta^i_1$, $y^i = B^{-1} \delta^i_2$, and $z^i = C^{-1} \delta^i_3$.
Subsequently, the Einstein field equations $G_{i j} = 8 \pi T_{i j}$ in natural units ($G=1$ and $c=1$)
can be expressed as follows:
\begin{eqnarray}
  & & \frac{\dot{A} \dot{B}}{A B} + \frac{\dot{A} \dot{C}}{A C} + \frac{\dot{B} \dot{C}}{B C} = 8 \pi \, \rho \, , \label{efe-1} \\ & & \frac{\ddot{A}}{A} + \frac{\ddot{B}}{B} + \frac{\dot{A} \dot{B}}{A B} = - 8 \pi P_x \, ,  \label{efe-2} \\ & & \frac{\ddot{A}}{A} + \frac{\ddot{C}}{C} + \frac{\dot{A} \dot{C}}{A C} = - 8 \pi P_y \, ,  \label{efe-3} \\ & & \frac{\ddot{B}}{B} + \frac{\ddot{C}}{C} + \frac{\dot{B} \dot{C}}{B C} = - 8 \pi P_z \, , \label{efe-4}
\end{eqnarray}
where $G_{i j} = R_{i j} - \frac{1}{2} R g_{ij}$ is the Einstein tensor, $R_{i j}$ is the Ricci tensor, and $R$ is the Ricci scalar, which has the form
\begin{equation}
  R = 2 \left( \frac{\ddot{A}}{A} + \frac{\ddot{B}}{B} + \frac{\ddot{C}}{C} + \frac{\dot{A} \dot{B}}{A B}   + \frac{\dot{A} \dot{C}}{A C} + \frac{\dot{B} \dot{C}}{B C} \right) \, .
\end{equation}

A curvature scalar characterizes the spacetime curvature. One of the important curvature scalars is the Kretschmann scalar, defined by $K= R_{i j k l} R^{i j k l}$, which is a quadratic scalar invariant of the Riemann tensor $R_{i j k l}$. It is crucial for measuring curvature in a vacuum.
This curvature scalar also characterizes the spacetime curvature of a realistic rotating black hole, allowing us to mathematically perceive the black hole. Additionally, we consider an important curvature scalar, the so-called Gauss--Bonnet (GB) invariant $\mathcal{G}$, defined as $\mathcal{G} = R^2 -4 R_{i j} R^{i j} + R_{i j k l} R^{i j k l}$. Apart from emerging in the context of defining quantum fields in curved spacetimes, the GB invariant $\mathcal{G}$ can encapsulate all the curvature information stemming from the Riemann tensor in dynamical equations. For the Bianchi I spacetime, both the $K$ and $\mathcal{G}$ take the following forms
\begin{eqnarray}
  & & K = 4 \left[ \frac{\ddot{A}^2}{A^2} + \frac{\ddot{B}^2}{B^2} + \frac{\ddot{C}^2}{C^2} + \frac{(\dot{A} \dot{B})^2}{(A B)^2}   + \frac{( \dot{A} \dot{C})^2}{(A C)^2} + \frac{(\dot{B} \dot{C})^2}{(B C)^2} \right] \, , \\ & & \mathcal{G} = 8 \left( \frac{\ddot{A}}{A} + \frac{\ddot{B}}{B} + \frac{\ddot{C}}{C} \right) \, .
\end{eqnarray}

In the previous section, we investigated the power law form of metric functions, such as $A(t) = t^L$, $B(t)=t^p$, and $C(t)=t^q$  in specific cases. With this form of the Bianchi I metric, analysis of the field Equations \eqref{efe-1} to \eqref{efe-3} reveals the expressions for the physical variables
$\rho, P_x, P_y$ and $P_z$ to be
\begin{eqnarray}
  & & \rho =  \frac{( L p + L q + p q)}{ 8 \pi \, t^2} \, , \qquad  P_x = - \frac{( p^2 + p q + q^2 - p - q)}{ 8 \pi \, t^2} \, , \label{rho-Px-power}  \\ & &  P_y = - \frac{( L^2 + L q + q^2 - L - q)}{8 \pi \, t^2} \, , \qquad P_z = - \frac{( L^2 + L p + p^2 - L - p)}{ 8 \pi \, t^2} \, , \label{Py-Pz-power}
\end{eqnarray}
As a result, the directional EoS parameters $w_x, w_y$, and $w_z$ along the $x, y$, and $z$ axes, respectively, are determined as follows:
\begin{eqnarray}
   & w_x = \frac{p + q - p^2 - p q - q^2 }{ L p + L q + p q } \, , \, w_y = \frac{L + q - L^2 - L q - q^2 }{ L p + L q + p q } \, , \, w_z = \frac{L + p - L^2 - L p - p^2 }{ L p + L q + p q } \, . \qquad
\end{eqnarray}
Subsequently, the average EoS parameter $w$ is expressed as
\begin{equation}
  w =  1 + \frac{2}{3} \frac{( L + p + q) (1 - L - p - q)}{ L p + L q + p q}  \, ,
\end{equation}
where $L p + L q + p q \neq 0$. Finally, the Ricci, Kretschmann, and the GB scalars for the power law form of metric coefficients are obtained as follows:
\begin{eqnarray}
  & & R = \frac{4}{t^2} \left( L^2 + p^2 + q^2 + L p + L q + pq - L -p - q \right) \, , \quad \mathcal{G} = \frac{ 8 L  p \, q \, ( L + p + q - 3)}{t^4} \, ,   \\ & & K = \frac{8}{t^4} \left[ L^4 + p^4 + q^4 - 2 ( L^3 + p^3 + q^3 ) + L^2 p^2 + L^2 q^2 + p^2 q^2 + L^2 + p^2 + q^2 \right] \, .
\end{eqnarray}
If $L p + L q + p q = 0$, both $p$ and $q$ should be zero, and there is no need for vanishing of the power $L$. This means that the energy density $\rho$ is zero, while the anisotropic pressures in $x, y$ and $z$ directions are $P_x = 0$, $P_y = L (1 - L)/(8 \pi t^2)$ and $P_z = P_y$, respectively.
In theoretical physics and cosmology, scenarios where the energy density vanishes (or is extremely low) while the pressure is different from zero are less common. However, there are still certain contexts where this can occur, often involving exotic fields or conditions. In certain models of the early universe, there is a concept known as the {\it false vacuum}. The vacuum state of a field can be thought of as the state of lowest energy. In a false vacuum, the energy density is very close to zero, but the field is in a metastable state rather than the true vacuum state.

When considering a perfect fluid source ($P_x = P_y = P_z = P$), Equations \eqref{rho-Px-power} to \eqref{Py-Pz-power} imply the following constraint relations:
\begin{equation}
  L ( L-1) - p (p-1) + q (L - p) = 0 \, , \qquad L (L-1) - q (q-1) + p (L-q) = 0 \, . \label{constr-1}
\end{equation}
For cases where $L= p = q$, as seen in cases (ii.2) and (ii.3), these constraint equations are precisely satisfied. Consequently, the physical quantities $\rho$ and $P$ take the forms
\begin{equation}
  \rho = \frac{3 q^2}{8 \pi \, t^2} \, , \qquad P = \frac{q ( 2 - 3 q)}{8 \pi \, t^2} \,  ,
\end{equation}
yielding the EoS parameter $w = (2 - 3 q)/(3 q)$ where $q \neq 0$. So, dark energy occurs when $q > 1$, which mentions that a well-known example of dark energy ($w=-1$), the cosmological constant, is not  possible for the power law Bianchi I spacetime with a perfect fluid. In addition, we have dust fluid if $q = 1/3$, and a stiff fluid if $q = 1/2$. Additionally, the curvature scalars
$R, K$ and $\mathcal{G}$ for cases (ii.1) and (ii.3) are given by
\begin{eqnarray}
  & & R = \frac{6 q (2 q -1)}{t^2} \, , \qquad K = \frac{12\, q^2 ( 2 q^2 - 2 q +1)}{t^4} \, , \qquad \mathcal{G} = \frac{24 q^3 (q-1)}{t^4} \, .
\end{eqnarray}
Moreover, when $L = p = q = 2/3$, the spacetime corresponds to the Einstein--de Sitter model, a solution of field Equations \eqref{efe-1} to \eqref{efe-3} with $P=0$ and $\rho = 1/(6 \pi \, t^2)$, i.e., $w=0$ (representing a dust fluid).

Kasner spacetimes, non-trivial Bianchi I solutions satisfying Einstein's vacuum equations
($R_{i j} = 0$, or $\rho = 0$ and $P = 0$), are characterized using the Kasner metric. In this metric, the three constant parameters $L, p$, and $q$ are termed the Kasner components. These components adhere to the following relations:
\begin{equation}
  L + p + q = 1 \, , \qquad L^2 + p^2 + q^2 = 1 \, , \label{constr-kasner}
\end{equation}
resulting in $ L p + L q + p q = 0$. Kasner spacetimes exhibit spatial anisotropy, potentially expanding along one direction while contracting along another. In the Kasner scenario, the three parameters cannot all be equal. Thus, at least one of $L, p, q$ must satisfy $0 \leq L, p, q \leq 1$. If any of the parameters $L, p$, or $q$ equals one, the other two must be zero, resulting in a flat Kasner spacetime. Conversely, if $L, p, q \neq 1$ are all not equal to one, indicating all are non-zero, the Kasner spacetime is non-flat \cite{kramer}.

%5
\section{Discussions and Concluding Remarks}
\label{sec:5}

Though the standard procedure for determining Lie and Noether symmetry generators can be cumbersome, it is feasible to consider reduction or obtaining conservation laws through Noether's theorem. This paper delved into the geometric nature of the Klein--Gordon/wave equation within the framework of the Lagrangian for the Bianchi I spacetime. We demonstrated that computing Noether symmetries of the first-order Lagrangian \eqref{lagr-2} for the Klein--Gordon Equation \eqref{kg-eq-2} is simplified to solving a set of differential conditions outlined by \eqref{ngs-cond-3}.
For the Bianchi I spacetime, employing this method to find Noether symmetries yields solutions encompassing scenarios where $F(t,x,y,z,\psi)= U_0$ (let us say, $U_0 = 0$ for convenience), reducing Equation \eqref{kg-eq-2} to the wave equation. Conversely, when $F (t,x,y,z,\psi) \neq {\rm const.}$, Equation \eqref{kg-eq-2} represents the Klein--Gordon equation. Exploring various functional forms of metric coefficients, we derived exact solutions of the Noether symmetry Equations \eqref{ngs-cond-3} for the Bianchi I spacetime.

In case (i), for $F = U_0 + U_1 \psi^2 + \frac{1}{2} U_2^2 \psi^2$, we obtained the Noether symmetry generators and gauge vector components for arbitrary metric coefficients. Additionally, we constructed conserved vector fields corresponding to these Noether symmetries. Selected solutions of Equation \eqref{ode-case-i} are presented in Table \ref{Tab1} for specific trigonometric or hyperbolic metric functions other than power-law forms, involving the unknown function $Y(t)$ in the components of Noether symmetry generators.

In case (ii), utilizing power-law forms of metric coefficients, we derived Noether symmetries in terms of constant powers $L, p$, and $q$. In Table \ref{Tab2}, a set of nontrivial solutions of Equation \eqref{ode-case-i1-1}, which solve $Y(t)$ akin to Equation \eqref{ode-case-i}, is provided for a specific class of Bianchi I metrics. In subcases of (ii), particularly (ii.1), (ii.2), and (ii.3), we discovered special values of powers $L, p$, and $q$, resulting in various dimensions of Noether symmetry groups. Additionally, Einstein field equations incorporating an imperfect fluid were derived for arbitrary metric functions within the context of Bianchi I spacetime and particularly studied for the power-law form of that geometry.

Let us briefly discuss a comparison and contrast between this study and some related references.
In \cite{ptm}, the study focused on the Lie symmetries applied to the Klein--Gordon and wave equations within Bianchi I spacetime in terms of  the potential function $V(x^i)$. They considered the function $F$ as $F = V(x^i) \, \psi^2 /2$, which gives $G (x^i, \psi) = - V(x^i) \, \psi$, specifically for the Klein--Gordon equation. In this context, the investigation in \cite{pt} revealed that for $F = V(x^i) \, \psi^2 /2$, the Noether symmetry component of the scalar field $\psi$ is \linebreak $\Phi = -2 \, \sigma \, \psi + b(x^k)$, with $b(x^k)$ being a solution of (3). Then, it follows from the Noether symmetry condition \eqref{cond-potential-1} that the equation $A^i_{,i} = 0$ and the relationship
\begin{equation}
  V_{,i} \xi^i + 2 \, \sigma \, V - 2 \, \square \, \sigma = 0 \, , \label{cond-potential-2}
\end{equation}
were established, incorporating the potential function $V(x^i)$ and the conformal factor $\sigma (x^i)$.
In contrast, our study directly delved into the Noether symmetry Equations \eqref{ngs-cond-3} employing an arbitrary function $F(x^i,\psi)$, bypassing the need to solve condition \eqref{cond-potential-2} to find potentials $V(x^i)$ concerning point symmetries generated by the spacetime symmetries of Bianchi I metric \eqref{b1}. We obtained solutions for these equations, including components of Noether symmetry generators, employing specific solvable forms of the function $F$ within Bianchi I spacetime. Consequently, we not only acquired conserved Noether currents but also derived non-trivial gauge vectors for each Noether symmetry.
Moreover, in the paper \cite{p2023}, the author presented solutions for the classification problem concerning the Klein--Gordon equation with a non-constant potential function $V(x^i)$ within the framework of Bianchi I, Bianchi III, and Bianchi V spacetimes. This investigation showcased that the Klein--Gordon equation admits non-trivial Lie symmetries derived from conformal algebra. However, the metric tensor provided for Bianchi I spacetime in that reference, given as
$$g_{i j} = {\rm diag} \left[ -C^2 (t), e^{-\frac{2}{c} t} C^2 (t), e^{-\frac{2 \alpha_1}{c} t} C^2 (t), C^2 (t) \right],$$, is less general compared to the one employed in our study, described by \eqref{b1}.
This study investigated the Klein--Gordon equation for Bianchi I spacetime, employing the Noether symmetry approach as a means to derive conservation laws. The novelty of this research lies in utilizing the Noether symmetry method to discern the symmetries inherent in the Lagrangian, thereby establishing the associated conservation laws for the Klein--Gordon equation of Bianchi I spacetime.

It is widely recognized that Lie point symmetries serve as a tool for reducing the Klein--Gordon/wave equation, thereby identifying corresponding invariant solutions. The reduction based on the Lie invariants of the symmetry vector leads to a solution of the Klein--Gordon/wave equation, coupled with a constraint equation related to an arbitrary integration function within the solution. In the paper \cite{abdul2017}, invariant solutions of the wave equation on Bianchi I spacetime using the Lie symmetry method were investigated. In our study, we used the power-law form of metric coefficients outlined in this reference. However, we extended the investigation by exploring both the Noether symmetries of the wave and Klein--Gordon equations, obtaining conserved quantities for each Noether symmetry. Considering that every Noether symmetry should be a Lie symmetry, we aimed to utilize Noether symmetries for reducing both the wave equation and Klein--Gordon equation. Meanwhile, the common Lie symmetry vector for Bianchi I spacetime is $\psi \partial \psi$ \cite{ptm}, which is not a Noether symmetry vector. Based on this observation, we inferred that the scalar field $\psi$, representing the solution of the Klein--Gordon equation, should be equal to the component $\Phi$ of the Noether symmetry generator ${\bf X}$; that is, $\psi = \Phi$. Further exploration of this relationship will be the focus of another paper.

%%%%%%%%%%%%%%%%%%%%%%%%%%%%%%%%%%%%%%%%%%%%%%%%%%%%%%%%%%%%%%%%%%%%%%%%%%%%%

% \section*{Acknowledgements}

% AQ acknowledges a research grant from the Higher Education Commission of Pakistan.

% \reftitle{References}
% \begin{thebibliography}{10}

%======================================================================

\end{document}